\documentclass[12pt]{article}
\usepackage{feynmp}
\usepackage{fullpage}
\newcommand {\be} {\begin{equation}}
\newcommand{\ee} {\end{equation}}
\newcommand{\bea}{\begin{eqnarray}}
\newcommand{\eea}{\end{eqnarray}}
\newcommand{\bean}{\begin{eqnarray*}}
\newcommand{\eean}{\end{eqnarray*}}
\newcommand{\noi}{\noindent}
\newcommand{\tn}{\textnormal}
\newcommand{\bi}{\bibitem}
\begin{document}

\begin{titlepage}

\rightline{Alberta Thy 04-01 }
\rightline{hep-th/0102022}
\rightline{January 2001}

\vskip .3in 

\begin{center}
{\large{\bf Noncommutative Linear Sigma Models}}
\end{center}

\vskip .2in

\begin{center}
Bruce A. Campbell and Kirk Kaminsky\\ 

\vskip .1in

{\it Department of Physics, University of 
Alberta}  \\ {\it  Edmonton, Alberta, Canada T6G 2J1} \\
\end{center}

\vskip .2in

\begin{abstract}
We examine noncommutative linear sigma models with $U(N)$ global 
symmetry groups at the one-loop quantum level, and contrast the 
results with our previous study of the noncommutative $O(N)$ linear 
sigma models where we have shown that Nambu-Goldstone symmetry 
realization is inconsistent with continuum renormalization.  Specifically we 
find no violation of Goldstone's theorem at one-loop for the $U(N)$ models 
with the quartic term ordering consistent with possible noncommutative 
gauging of the model.  The difference is due to terms involving 
noncommutative commutator interactions, which vanish in the commutative 
limit.  We also examine the $U(2)$, and $O(4)$ linear sigma models with 
matter in the adjoint representation, and find that the former is
consistent with Goldstone's theorem at one-loop if we include 
only trace invariants consistent with possible noncommutative gauging of 
the model, while the latter exhibits violations of Goldstone's theorem of 
the kind seen in the fundamental of $O(N)$ for $N>2$. 
\end{abstract}

\vskip .1in

\end{titlepage}

\newpage
\section{Introduction}

Recently field theories on noncommutative spacetime backgrounds have been the 
subject of intense scrutiny \cite{general}.  Part of this motivation stems from the 
fact that noncommutative $U(N)$ gauge theories arise on D-branes in the 
presence of a constant NS-NS B-field background, in the zero-slope, field 
theoretic limit of string theory \cite{sch},\cite{sw}.  A second motivation,
independent of string theory, is the question of whether the world we live
in is based on a noncommutative spacetime.  In order to construct realistic
models of particle physics on noncommutative spacetimes, one needs to be
sure that noncommutative theories preserve the features that underlie the
standard model, including perturbative renormalizability in the presence
of spontaneous symmetry breaking\cite{ssb1},\cite{ssb2}.

The general scheme for defining field theories with the noncommutative 
spacetime structure defined by $[ \hat{x}^{\mu}, \hat{x}^{\nu} ] = i 
\theta^{\mu\nu}$, $\theta^{\mu\nu}$ real, constant and antisymmetric, 
is to invoke Weyl-Moyal correspondence.  This has the effect of 
replacing the underlying noncommutative spacetime with a commutative 
spacetime at the expense of replacing the ordinary pointwise product 
of spacetime dependent functions with an infinitely nonlocal star 
product.  The induced momentum space Feynman rules for interaction
vertices associated with a given field theory then involve momentum-dependent 
phases, which generically split a graph (at least at one-loop) into planar 
and nonplanar parts.  The former are identical to the usual commutative graphs 
(up to a total phase depending only on the external momenta, and a 
combinatorial reweighting), and in particular possess the usual divergence 
structure associated with a commutative quantum field theory.  The latter, 
nonplanar components are explicitly finite (at least at one-loop) because of 
oscillatory damping due to the phases, and replace an ultraviolet divergence 
with an infrared divergence in the external momenta \cite{mrs},\cite{rs}.

Superficially, as a consequence of the finiteness of nonplanar 
graphs, and of the similar divergence structure of the planar graphs, one 
might conclude that the renormalization of noncommutative 
field theories proceeds as in the commutative theory, because the 
counterterm structure is formally the same.  However, as is well-known, 
the renormalization of spontaneously broken theories, with either 
underlying global or gauge symmetries, is more subtle because the 
number of counterterm vertices exceeds the number of renormalization 
parameters.  As a result, the renormalizability of (commutative) 
spontaneously broken theories hinges in general on intricate graphwise 
cancellations \cite{ssb1}, \cite{ssb2} order by order in 
perturbation theory.  Thus it is of obvious interest to examine 
whether or not these cancellations persist in noncommutative field 
theories.

In a previous paper \cite{us} we studied the spontaneous symmetry breaking of 
a global $O(N)$ symmetry in the noncommutative deformation of the linear 
sigma model with scalars in the fundamental representation.  We found that 
one-point tadpoles of the sigma at 
one-loop were insensitive to the noncommutativity because no external 
momentum flows into the trilinear tadpole vertex.  Thus the 
one-point sigma counterterm is identical to the one in the commutative limit, 
which in turn fixes the pion mass counterterm to be the same as its 
commutative limit.  On the other hand, the planar components of the 
1PI graphs contributing to the one-loop pion (inverse) propagator 
renormalization are re-weighted with respect to the corresponding 
commutative graphs.  As a consequence, there is an unavoidable UV cutoff 
dependence (for nonzero external momentum) {\it after} renormalization, 
signalling the nonexistence of a continuum limit, and noncommuting 
UV ($\Lambda_{UV} \rightarrow \infty$) and IR ($p\rightarrow 0$) limits. 
Specifically we found that the sum of the 1PI graphs and the 
counterterm contributing to the pion mass renormalization yielded 
quadratic and logarithmic UV cutoff $\Lambda$ dependences as:
\bea
\sum_{1-loop} &=& {\lambda \delta^{ij}\over 16\pi^2} \left\{ \bigg[ N (1-f) 
+ f \bigg] \Lambda^{2} \left( 1- {1\over 1 + \Lambda^{2} (p \circ p)} 
\right) - \bigg(2-f \bigg) m_{\sigma}^{2} \log(1+\Lambda^{2} (p\circ 
p)) \right\} \nonumber \\ & & 
+ \tn{finite} * p^{2} 
\eea
respectively.  Here $N$ is the dimension of the fundamental of $O(N)$, 
$p \circ q \equiv - p_{\mu} \theta^{2}_{\mu\nu} q_{\nu} /4$, and $f$ 
takes into account the two possible quartic orderings for $\pi\pi\pi\pi$ 
and $\pi\pi\sigma\sigma$ terms:
\bea
& & {\lambda \over 4} f (\pi^{k} \ast \pi^{k}) \ast ( \pi^{l} \ast \pi^{l}) 
+ {\lambda \over 4} (1-f) (\pi^{k} \ast \pi^{l}) \ast ( \pi^{k} \ast \pi^{l})
\nonumber \\ & & + {\lambda\over 2} f ( \pi^{k} \ast \pi^{k} ) \ast \sigma \ast \sigma + 
{\lambda\over 2} \left( 1-f \right) (\pi^{k} \ast \sigma) \ast 
(\pi^{k} \ast \sigma) \subset {\cal L}
\eea
For nonzero $\theta$ and $p$, the only circumstance under which we can take 
the continuum limit is when $f=2$ {\it and} $N=2$, where both 
logarithmic and quadratic dependences on $\Lambda$ vanish.  This 
corresponds to the {\it Abelian} $O(2)$ model, and if written in 
terms of a complex scalar $\phi$ corresponds precisely to the ordering 
\footnote{The interested reader
may verify that the $N=2$, $f=2$ case of our general analysis, is precisely the
global limit of the Abelian $U(1)\equiv O(2)$ model subsequently considered 
in \cite{p} as
can be easily seen by comparing Lagrangian densities, and Feynman rules; 
the conclusions of reference \cite{p} agree with our previous
analysis \cite{us}.} $\phi^{*} \ast \phi \ast \phi^{*} \ast \phi$.  Otherwise, the conditions $f=N/(N-1)$ and $f=2$ 
required for the cancellations of quadratic and logarithmic 
dependences on $\Lambda$ respectively, cannot be simultaneously satisfied, 
Goldstone's theorem fails at the one-loop level, and the continuum limit 
of the model fails to exist.  Thus for general $N>2$ and for {\it all} 
possible orderings consistent with the global $O(N)$ symmetry, the 
noncommutative $O(N)$ linear sigma model does not exist in the continuum limit.

{\it Prima facie}, this incompatibility of continuum renormalizability with
spontaneous symmetry breaking for $O(N)$ linear sigma models appears to
present severe difficulties for attempts to make realistic models of particle
physics on noncommutative spacetimes.  First, it is clear that models with
spontaneously broken {\it gauge} symmetries must have consistent spontaneously
broken global limits (as the gauge couplings vanish); the absence of such a
global limit with spontaneous symmetry breaking would preclude its
subsequent gauging (at least perturbatively).  Second, the standard model
of the fundamental interactions (and unified theories which encompass it) 
depends, for electroweak symmetry breaking, on a complex Higgs doublet.  As
is well known, resolved into real components, the purely scalar sector of the
standard model is $O(4)$ invariant (and not just $SU(2)\times U(1)$ invariant),
with the real components in the fundamental representation; our previous
results then appear to preclude noncommutative deformation of the standard 
model.  We will argue below that this is {\it not} necessarily the case.  
In particular, noncommutative theories with $N$ complex scalars, $\Phi^{i}$
($i=1..N$,$N>1$), and with $U(N)$ invariant self-interactions, are {\it not} 
invariant under an $O(2N)$ symmetry acting on their real components, due to 
purely noncommutative commutator interactions arising from the 
noncommutativity of
the spacetime.  Thus we will first undertake an analysis for the case of 
a $U(N)$ symmetry group with the scalars in the fundamental representation, 
choosing the quartic invariant $\Phi^{\dagger} \Phi \Phi^{\dagger} \Phi$.  
The spontaneous breaking of this group to $U(N-1)$ 
leaves $N-1$ complex pions, and one real pion.  We find that the 
one-loop 1PI graphs contributing to the mass renormalization of the complex pions, 
like the one-point tadpoles, do not see the noncommutativity at this order, and so 
Goldstone's theorem holds.  The 1PI one-loop graphs contributing to 
the mass renormalization of the real pion (which arises through the 
breaking of the $U(1)\equiv O(2)$ subgroup of the $U(N)$), are split into 
divergent planar, and finite nonplanar pieces in such a way that Goldstone's 
theorem holds at one-loop.  The essential difference between the $U(N)$ 
models and the corresponding $O(2N)$ models ($N>1$) is the presence of the 
purely noncommutative commutator interactions in the former.

We will also begin to explore how our present, and previous \cite{us}, results
might depend on the scalar field representation responsible for spontaneously 
breaking the symmetry.  In particular, we consider both an $O(4)$ and a 
$U(2)$  model, with scalars in the adjoint representation, to see if our
previous results depended on our scalars being in the respective 
fundamentals.  For the $U(2)$ model with matter in the adjoint 
representation, we will find that Goldstone's theorem holds if we include 
only interactions involving a single trace operator, which we will in 
turn demonstrate are the only ones consistent with noncommutative 
gauge invariance in the case that we gauge the $U(2)$ symmetry.  In this 
model Goldstone's theorem holds due to a 
notable cancellation of a purely noncommutative graph involving couplings 
to the $U(1)$ component of the field.  For the $O(4)$ model 
with matter in the adjoint, we find violations of Goldstone's 
theorem at one-loop of the type found in the $O(N)$ fundamental 
representation studied in \cite{us}.  Finally we discuss the implications of 
these results for model building, and comment on the nature of the IR 
divergences found by \cite{rs} in the context of noncommutative theories with 
matter in the adjoint representation.


\section{NC $U(N)$ Linear Sigma model: Fundamental Rep}

In this section we examine Goldstone's theorem in the noncommutative 
deformation of the linear sigma model with a global $U(N)$ symmetry group,
and contrast the results with our previously discovered violations of 
Goldstone's theorem in the $O(N)$ linear sigma model.

The noncommutative $U(N)$ linear sigma model is defined by the 
Lagrangian density given by
\be
{\cal L} = \partial_{\mu} \Phi^{\dagger} \ast \partial^{\mu} \Phi +
\mu^{2} \Phi^{\dagger} \ast \Phi - \lambda \Phi^{\dagger} \ast \Phi 
\ast \Phi^{\dagger} \ast \Phi 
\label{U(N)-Lag-1}
\ee
where $\Phi$ is an N-vector of {\it complex} fields 
$\phi_{i}$ ($i=1..N$), where the star product is defined as usual by
$f(x) * g(x) = \exp{(i \theta^{\mu\nu} \partial_{\mu}^{y} 
\partial_{\nu}^{z})} f(y) g(z) |_{y,z\rightarrow x}$, and where we 
have included the star ordering of the quartic term consistent with 
noncommutative gauge invariance of a possible gauging of the model 
(see below).  For $\mu^{2}>0$, the symmetry is 
spontaneously broken to $U(N-1)$.  Throughout the 
remainder of this paper, we will consider only translationally invariant   
vacua\footnote{As Gubser and Sondhi have argued \cite{gs}, more exotic vacua 
such as stripe phases are possible in noncommutative theories.}.  
By an $SU(N)$ transformation, we can rotate the resulting VEV into the last 
field of $\Phi$, and by a $U(1)$ rotation we can identify this VEV 
with a constant shift, $a$, in the real part of this field.  Thus we define 
$\pi_{i} = \phi_{i}$ for $i=1..N-1$, while $\phi_{N} = (\sigma + a + 
i\pi_{0})/\sqrt{2}$; there are $N-1$ complex Goldstone bosons, and one 
real Goldstone mode.  The minimization of the potential for this 
configuration implies:
\be
V(a) = \frac{\mu^{2}}{2} a^{2} - \frac{\lambda}{4} a^{4} \longrightarrow 
a^{2}  = \frac{\mu^{2}}{\lambda}
\ee
Writing (\ref{U(N)-Lag-1}) in terms of these variables yields:
\bea
{\cal L} &=&  \frac{1}{2} (\partial_{\mu} \sigma)^{2} + \frac{1}{2}
(\partial_{\mu} \pi_{0})^{2} + \partial_{\mu} \pi_{i}^{*} 
\partial^{\mu} \pi_{i} - \frac{1}{2}(2\mu^{2})\sigma^{2} - \lambda a 
\sigma^{3} \nonumber \\ & & -\lambda a \pi_{0}^{2} \sigma - 2 \lambda 
a \pi_{i}^{*} \pi_{i} \sigma - \lambda \pi_{i}^{*} \pi_{i} \pi_{j}^{*}
\pi_{j} - \frac{\lambda}{4}(\sigma^{4}+\pi_{0}^{4}) \nonumber \\ & & - \lambda 
\pi_{0}^{2} \sigma^{2} + \frac{\lambda}{2} \sigma \pi_{0} \sigma 
\pi_{0} - \lambda \pi_{i}^{*} \pi_{i} (\sigma^{2} + \pi_{0}^{2}) 
- \lambda \pi_{i}^{*} \pi_{i} [ \sigma,\pi_{0}]    
\eea
For notational brevity all star products will be suppressed henceforth, 
unless there is danger of confusion.  Furthermore, we will implicitly 
use the identity
\be
\int A_{1} * \ldots * A_{n} = \int A_{\sigma(1)} * \ldots * 
A_{\sigma(n)}   
\ee
(where $\{\sigma(1)\ldots\sigma(n)\}$ represents any cyclic permutation 
of $\{1\ldots n\}$), with the understanding that all Lagrangian density 
terms sit under a spacetime integral.  This identity means that 
quadratic terms in the action, and hence propagators, are identical to their 
commutative counterparts.  

To simplify the discussion relative to that occurring in \cite{us}, we will 
not {\it a priori} impose the vanishing of the tadpole as a 
renormalization condition.  Instead we will include the one-point 
tadpole contributions, and their counterterm directly in 
calculating the mass renormalization of the pion.  In this completely 
equivalent language, the two counterterms present cancel each other, 
up to the wavefunction renormalization, so the sum of the one-particle 
irreducible (1PI) graphs and the one-point tadpole insertions must be 
automatically finite up to wavefunction renormalization (and for 
Goldstone's theorem to hold at one-loop, must vanish in the 
$p\rightarrow 0$ limit)\cite{cl}.  Furthermore, to exhibit the essentially
algebraic nature of the result, we will expand the non-phase 
part of the integrands about zero-external momentum, in the
cases where there are two propagators in the loop using the Taylor expansion
\be
\frac{1}{k^2 [(p+k)^2 - m^2]} = \frac{1}{k^2 (k^2-m^2)} - p_{\mu} \frac{2 
k^{\mu}} {k^2 [k^2-m^2]^2} + \ldots
\ee
and then note that the $p$-dependent terms yield finite loop-momentum
integrals (for all $p$), and vanish as $p\rightarrow 0$.  We then define the 
momentum integrals
\be
I(m^{2}) = \int \frac{d^{4}k}{(2\pi)^{4}} \frac{1}{k^{2}-m^{2}} 
\hspace{5pt}, \hspace{5pt}
I_{\theta,p}(m^{2}) =  \int \frac{d^{4}k}{(2\pi)^{4}} \frac{\cos(k\times p)}
{k^{2}-m^{2}} = \int \frac{d^{4}k}{(2\pi)^{4}} \frac{\tn{e}^{i k\times p}}
{k^{2}-m^{2}}  
\ee
where $k\times p = k_{\mu} \theta^{\mu\nu} p_{\nu}$.

The vertices for the theory are listed in the Appendix, and the 
propagators are the usual ones.  Dashes denote complex pions, dots the
real pion associated with the sigma, and solid lines denote the sigma.  The 
1PI one-loop graphs contributing to the mass renormalization of the 
$N-1$ complex pions are
\begin{fmffile}{global1}
 \bea
   \parbox{120pt}{\begin{fmfgraph*}(120,60)
     \fmfleft{w}
     \fmfright{e}
     \fmf{dashes_arrow,label=$p$}{w,c,e}
     \fmffreeze
     \fmftop{n}
     \fmf{dashes,left,label=$k$}{c,n}
     \fmf{dashes,right}{c,n}
     \fmfdot{c}
     \fmfv{label=$\pi_{i}$,label.angle=60}{w}
     \fmfv{label=$\pi_{j}^{*}$,label.angle=120}{e}    
    \end{fmfgraph*}} \equiv (a) ; \hspace{10pt} 
   \parbox{120pt}{\begin{fmfgraph*}(120,60)
     \fmfleft{w}
     \fmfright{e}
     \fmf{dashes_arrow,label=$p$}{w,c,e}
     \fmffreeze
     \fmftop{n}
     \fmf{dots,left,label=$k$}{c,n}
     \fmf{dots,right}{c,n}
     \fmfdot{c}
     \fmfv{label=$\pi_{i}$,label.angle=60}{w}
     \fmfv{label=$\pi_{j}^{*}$,label.angle=120}{e}    
    \end{fmfgraph*}} \equiv (b) \nonumber \\
   \parbox{120pt}{\begin{fmfgraph*}(120,60)
     \fmfleft{w}
     \fmfright{e}
     \fmf{dashes_arrow,label=$p$}{w,c,e}
     \fmffreeze
     \fmftop{n}
     \fmf{plain,left,label=$k$}{c,n}
     \fmf{plain,right}{c,n}
     \fmfdot{c}
     \fmfv{label=$\pi_{i}$,label.angle=60}{w}
     \fmfv{label=$\pi_{j}^{*}$,label.angle=120}{e}    
    \end{fmfgraph*}} \equiv (c) ; \hspace{10pt}
   \parbox{120pt}{\begin{fmfgraph*}(120,60)
    \fmfleft{w}
    \fmfright{e}
    \fmf{dashes_arrow,label=$p$}{w,cw}
    \fmf{dashes,right,tension=0.5,label=$k$}{ce,cw}
    \fmf{plain_arrow,right,tension=0.5,label=$k+p$}{cw,ce}
    \fmf{dashes}{ce,e}
    \fmfdot{cw,ce}
    \fmfv{label=$\pi_{i}$,label.angle=60}{w}
    \fmfv{label=$\pi_{j}^{*}$,label.angle=120}{e}    
   \end{fmfgraph*}} \equiv (d)   
 \eea

They are given respectively by
\bea
(a) &=& - 2i \lambda i \int \frac{d^{4}k}{(2\pi)^{4}} \frac{\delta_{kl} 
\left[ \delta^{ij}\delta^{kl} \tn{e}^{0} + \delta^{il}\delta^{jk} e^{0} 
\right]}{k^{2}} = 2 N \lambda \delta^{ij} I(0) \nonumber \\        
(b) &=& \frac{-2i\lambda i \delta^{ij}}{2} \int \frac{d^{4}k}{(2\pi)^{4}}
\frac{\tn{e}^{0} \cos(0)}{k^{2}} = \lambda \delta^{ij} I(0)   \nonumber \\
(c) &=& \frac{-2i\lambda i \delta^{ij}}{2} \int \frac{d^{4}k}{(2\pi)^{4}}
\frac{e^{0} \cos(0)}{k^{2}-2\mu^{2}} = \lambda \delta^{ij} I(2\mu^{2}) \nonumber \\
(d) &=& (-2i\lambda a)^{2} i^{2}\delta^{ik}\delta_{kl}\delta^{lj} 
\int \frac{d^{4}k}{(2\pi)^{4}} \frac{ \tn{e}^{-\frac{i}{2} (k\times p)} 
\tn{e}^{-\frac{i}{2} (-p\times -k)}}{[(p+k)^{2}-2\mu^{2}]k^{2}} \nonumber \\
&=& \frac{4 \lambda^{2} a^{2}}{2\mu^{2}} \delta^{ij} 
\int \frac{d^{4}k}{(2\pi)^{4}} \left[ \frac{1}{k^{2}- 2\mu^{2}}
- \frac{1}{k^{2}} \right] + \delta^{ij} A^{\mu} p_{\mu} \nonumber \\
&=& 2 \lambda \delta^{ij} \left[ I(2\mu^{2}) - I(0)\right] + 
\delta^{ij} A^{\mu} p_{\mu}    
\eea
where $A^{\mu}(p)$ is finite for all $p$.  
Evidently these 1PI graphs do not see the noncommutativity.  Meanwhile the 
one-point tadpoles insertions, as in \cite{us}, also do not see the 
noncommutativity at one-loop.  They are given by
\bea
   \parbox{100pt}{\begin{fmfgraph*}(100,60)
     \fmfleft{w}
     \fmfright{e}
     \fmf{dashes_arrow,label=$p$}{w,c,e}
     \fmffreeze
     \fmftop{n}
     \fmf{plain,tension=2.5}{c,cn}
     \fmf{dashes,right}{cn,n}
     \fmf{dashes,left}{cn,n}
     \fmfdot{c}
     \fmfdot{cn}
     \fmfv{label=$\pi_{i}$,label.angle=60}{w}
     \fmfv{label=$\pi_{j}^{*}$,label.angle=120}{e}    
    \end{fmfgraph*}} \equiv (e) ; 
   \parbox{100pt}{\begin{fmfgraph*}(100,60)
     \fmfleft{w}
     \fmfright{e}
     \fmf{dashes_arrow,label=$p$}{w,c,e}
     \fmffreeze
     \fmftop{n}
     \fmf{plain,tension=2.5}{c,cn}
     \fmf{plain,right}{cn,n}
     \fmf{plain,left}{cn,n}
     \fmfdot{c}
     \fmfdot{cn}
     \fmfv{label=$\pi_{i}$,label.angle=60}{w}
     \fmfv{label=$\pi_{j}^{*}$,label.angle=120}{e}    
    \end{fmfgraph*}} \equiv (f) ;
   \parbox{100pt}{\begin{fmfgraph*}(100,60)
     \fmfleft{w}
     \fmfright{e}
     \fmf{dashes_arrow,label=$p$}{w,c,e}
     \fmffreeze
     \fmftop{n}
     \fmf{plain,tension=2.5}{c,cn}
     \fmf{dots,right}{cn,n}
     \fmf{dots,left}{cn,n}
     \fmfdot{c}
     \fmfdot{cn}
     \fmfv{label=$\pi_{i}$,label.angle=60}{w}
     \fmfv{label=$\pi_{j}^{*}$,label.angle=120}{e}    
    \end{fmfgraph*}} \equiv (g)
\eea    
Their values are given by
\bea
(e) &=& (-2 i \lambda a \delta^{ij}) \frac{i}{-2\mu^{2}} (-2i\lambda 
a \delta^{kk}) i I(2\mu^{2}) = - 2(N-1) \lambda \delta^{ij} 
I(0) \nonumber \\  
(f) &=& (-2 i \lambda a \delta^{ij}) \frac{i}{-2\mu^{2}} (-6i\lambda 
a ) \frac{i}{2} I(2\mu^{2}) = - 3\lambda \delta^{ij} I(2\mu^{2}) 
\nonumber \\
(g) &=& (-2 i \lambda a \delta^{ij}) \frac{i}{-2\mu^{2}} (-2i\lambda 
a) \frac{i}{2} I(2\mu^{2}) = - \lambda \delta^{ij} I(0) 
\eea
where all noncommutative vertices manifestly collapse.

The sum of these seven graphs is equal to zero (modulo the finite 
term which itself vanishes as $p\rightarrow 0$), independently of a 
regulator, and of the noncommutativity, whence Goldstone's theorem 
holds at one-loop; the complex pions undergo no mass renormalization.  
Now consider the one-loop mass renormalization of $\pi_{0}$.
The 1PI graphs contributing are given by
 \bea
   \parbox{120pt}{\begin{fmfgraph*}(120,60)
     \fmfleft{w}
     \fmfright{e}
     \fmf{dots_arrow,label=$p$}{w,c,e}
     \fmffreeze
     \fmftop{n}
     \fmf{dots,left,label=$k$}{c,n}
     \fmf{dots,right}{c,n}
     \fmfdot{c}
     \fmfv{label=$\pi_{0}$,label.angle=60}{w}
     \fmfv{label=$\pi_{0}$,label.angle=120}{e}    
    \end{fmfgraph*}} \equiv (h) ; \hspace{10pt} 
   \parbox{120pt}{\begin{fmfgraph*}(120,60)
     \fmfleft{w}
     \fmfright{e}
     \fmf{dots_arrow,label=$p$}{w,c,e}
     \fmffreeze
     \fmftop{n}
     \fmf{plain,left,label=$k$}{c,n}
     \fmf{plain,right}{c,n}
     \fmfdot{c}
     \fmfv{label=$\pi_{0}$,label.angle=60}{w}
     \fmfv{label=$\pi_{0}$,label.angle=120}{e}    
    \end{fmfgraph*}} \equiv (i) \nonumber \\
   \parbox{120pt}{\begin{fmfgraph*}(120,60)
     \fmfleft{w}
     \fmfright{e}
     \fmf{dots_arrow,label=$p$}{w,c,e}
     \fmffreeze
     \fmftop{n}
     \fmf{dashes,left,label=$k$}{c,n}
     \fmf{dashes,right}{c,n}
     \fmfdot{c}
     \fmfv{label=$\pi_{0}$,label.angle=60}{w}
     \fmfv{label=$\pi_{0}$,label.angle=120}{e}    
    \end{fmfgraph*}} \equiv (j) ; \hspace{10pt}
   \parbox{120pt}{\begin{fmfgraph*}(120,60)
    \fmfleft{w}
    \fmfright{e}
    \fmf{dots_arrow,label=$p$}{w,cw}
    \fmf{dots,right,tension=0.5,label=$k$}{ce,cw}
    \fmf{plain_arrow,right,tension=0.5,label=$k+p$}{cw,ce}
    \fmf{dots}{ce,e}
    \fmfdot{cw,ce}
    \fmfv{label=$\pi_{0}$,label.angle=60}{w}
    \fmfv{label=$\pi_{0}$,label.angle=120}{e}    
   \end{fmfgraph*}} \equiv (k)   
 \eea
with values given by
\bea
(h) &=& \frac{-2i\lambda}{2} \int \frac{d^{4}k}{(2\pi)^{4}}
\frac{i \left[ 2 \cos^{2}(\frac{k\times p}{2}) + 1 \right]}{k^{2}}
= 2 \lambda I(0) + \lambda I_{\theta,p}(0) \nonumber \\
(i) &=& \frac{-2i \lambda}{2} \int \frac{d^{4}k}{(2\pi)^{4}}
\frac{i \left[ 2 \cos^{2}(0) - cos(k\times p) 
\right]}{k^{2}-2\mu^{2}} = 2\lambda I(2\mu^{2}) - \lambda 
I_{\theta,p}(2\mu^{2}) \nonumber \\      
(j) &=& -2 i \lambda \delta^{kk} \int \frac{d^{4}k}{(2\pi)^{4}}
\frac{ i \tn{e}^{0} \cos(0)}{k^{2}} = 2 (N-1) \lambda I(0) \nonumber \\  
(k) &=& (-2 i \lambda a)^{2} \int \frac{d^{4}k}{(2\pi)^{4}}
\frac{ i^{2} \cos^{2}(\frac{p\times k}{2})}{[ (p+k)^{2}-2\mu^{2}]k^{2}}
\nonumber \\
&=& \frac{4\lambda^{2}a^{2}}{2\mu^{2}} \int \frac{d^{4}k}{(2\pi)^{4}}
\frac{1}{2}(1+\cos(k\times p) \left[ \frac{1}{k^{2}- 2\mu^{2}}
- \frac{1}{k^{2}} \right] + B^{\mu}_{\theta} p_{\mu} \nonumber \\
&=& \lambda [ I(2\mu^{2}) - I(0)] + \lambda [ I_{\theta,p}(2\mu^{2}) -
I_{\theta,p}(0) ] + B^{\mu}_{\theta} p_{\mu}   
\eea
where $B^{\mu}_{\theta}(p)$ is finite for all $p$.  Evidently, the 
nonplanar contributions due to the noncommutativity cancel between 
these graphs.  For completeness, the one-point tadpole insertions are
\bea
   \parbox{100pt}{\begin{fmfgraph*}(100,60)
     \fmfleft{w}
     \fmfright{e}
     \fmf{dots_arrow,label=$p$}{w,c,e}
     \fmffreeze
     \fmftop{n}
     \fmf{plain,tension=2.5}{c,cn}
     \fmf{dashes,right}{cn,n}
     \fmf{dashes,left}{cn,n}
     \fmfdot{c}
     \fmfdot{cn}
     \fmfv{label=$\pi_{0}$,label.angle=60}{w}
     \fmfv{label=$\pi_{0}$,label.angle=120}{e}    
    \end{fmfgraph*}} \equiv (l) ; 
   \parbox{100pt}{\begin{fmfgraph*}(100,60)
     \fmfleft{w}
     \fmfright{e}
     \fmf{dots_arrow,label=$p$}{w,c,e}
     \fmffreeze
     \fmftop{n}
     \fmf{plain,tension=2.5}{c,cn}
     \fmf{plain,right}{cn,n}
     \fmf{plain,left}{cn,n}
     \fmfdot{c}
     \fmfdot{cn}
     \fmfv{label=$\pi_{0}$,label.angle=60}{w}
     \fmfv{label=$\pi_{0}$,label.angle=120}{e}    
    \end{fmfgraph*}} \equiv (m) ;
   \parbox{100pt}{\begin{fmfgraph*}(100,60)
     \fmfleft{w}
     \fmfright{e}
     \fmf{dots_arrow,label=$p$}{w,c,e}
     \fmffreeze
     \fmftop{n}
     \fmf{plain,tension=2.5}{c,cn}
     \fmf{dots,right}{cn,n}
     \fmf{dots,left}{cn,n}
     \fmfdot{c}
     \fmfdot{cn}
     \fmfv{label=$\pi_{0}$,label.angle=60}{w}
     \fmfv{label=$\pi_{0}$,label.angle=120}{e}    
    \end{fmfgraph*}} (n)
 \eea    
\end{fmffile}
with values given by
\bea
(l) &=& (-2i\lambda a)\frac{i}{-2\mu^{2}}(-2i\lambda a\delta^{kk})i 
I(0) = -2 (N-1)\lambda I(0) \nonumber \\
(m) &=& (-2i\lambda a)\frac{i}{-2\mu^{2}}(-6i\lambda a)\frac{i}{2} 
I(2\mu^{2}) = -3\lambda I(2\mu^{2}) \nonumber \\
(n) &=& (-2i\lambda a)\frac{i}{-2\mu^{2}}(-2i\lambda a)\frac{i}{2} 
I(0) = -\lambda I(0) \nonumber \\
\eea
The sum of these seven graphs also vanishes (again modulo the finite 
$p$ dependent term, which vanishes as $p\rightarrow 0$); so that 
Goldstone's theorem also holds for the neutral pion of this model.

Let us reflect on these results.  First, had we included the other 
ordering of the quartic term $\phi_{i}^{*} \ast \phi_{j}^{*} \ast
\phi_{i} \ast \phi_{j}$, we would again find violations of Goldstone's 
theorem of the type found in \cite{us}.  Secondly, we contrast these 
results with those of the general $O(N)$ model studied in \cite{us}, where
we showed violations of Goldstone's theorem at one-loop for {\it all} 
orderings consistent with the $O(N)$ global symmetry (except for the trivial 
Abelian case).  The difference here of course is that we are working with
a $U(N)$ group, which we now show exhibits crucial algebraic 
differences with the $O(N)$ case in noncommutative scalar theories.

Matter in the fundamental of $O(N)$ is described by a {\it real} N-vector
of fields, which we denote by $\Psi$.  As such, the invariant term $\Psi^{T} 
\ast \Psi$ merely is the sum of squares of the real components.  Then, the 
expansion of the quartic invariant can yield cross-terms only of the form 
$\psi_{i}\ast\psi_{i}\ast\psi_{j}\ast\psi_{j}$, or $\psi_{i}\ast\psi_{j}
\ast\psi_{i}\ast\psi_{j}$ corresponding to the two possible 
star product orderings of such an invariant.  Note that no more than 
two distinct fields can occur.   

On the other hand, the fundamental of $U(N)$ is described by a {\it complex} 
N-vector of fields, $\Phi$.  Now however, the quadratic invariant 
$\Phi^{\dagger} * \Phi$, written in terms of real fields picks up the 
{\it commutator} of each field's real part with its imaginary part due to the 
noncommutativity since
\be
( R - i I) (R + i I) = R^{2} + I^{2} + i [ R,I]  
\ee
While such commutators in the quadratic term vanish 
when integrated over spacetime, the quartic invariant now picks up 
products of such commutators with other fields or commutators which, 
for $N>1$, constitute new interactions between real components 
of two complex $\phi$'s, not present in, and incompatible with, the 
$O(2N)$ symmetry. 

Let us make this argument manifest.  Expanding the quartic term in 
the $U(N)$ theory in terms of its real components yields
\bea
\phi^{*}_{i} \phi_{i} \phi^{*}_{j} \phi_{j} &=&  
(\phi_{iR} - i \phi_{iI}) (\phi_{iR} + i \phi_{iI}) (\phi_{jR} - i 
\phi_{jI}) (\phi_{jR} + i \phi_{jI}) \nonumber \\
&=& \phi_{iR}^{2} \phi_{jR}^{2} + \phi_{iR}^{2}\phi_{jI}^{2} + 
\phi_{iI}^{2} \phi_{jR}^{2} + \phi_{iI}^{2}\phi_{jI}^{2} 
- [ \phi_{iR}, \phi_{iI}] [\phi_{jR}, \phi_{jI}] 
\nonumber \\
& & + i (\phi_{iR}^{2} + \phi_{iI}^{2}) [\phi_{jR},\phi_{jI}]
+ i (\phi_{jR}^{2} + \phi_{jI}^{2}) [\phi_{iR},\phi_{iI}]
\eea
We note that for $i\neq j$ (which can occur for $N>1$), the presence of
interactions (those involving three or four distinct real fields) which 
{\it cannot} occur in the $O(2N)$ case by the general argument above.  We 
emphasize this is a purely noncommutative effect\footnote{For 
$i=j$ (or $N=1$), the last two terms vanish under the spacetime integral, 
and the product of commutators merely induces the orderings of the 
$O(2)$ model studied in \cite{us} with $f=2$.}.  The presence of these 
extra, purely noncommutative interactions is responsible for the the differing 
behaviour of the spontaneously broken phase at the quantum level for 
these models.

To conclude, we have found that one cannot in general spontaneously break a 
fundamental representation NC $O(N)$ linear sigma model, 
while one can break a fundamental representation NC $U(N)$ linear sigma model 
for the noncommutative gauge invariant quartic ordering.  This latter theme is 
one that will arise again in a more dramatic fashion in the adjoint 
representation model to which we now turn.


\section{NC $U(2)$ Sigma Model: Adjoint Representation}

We now examine the status of Goldstone's theorem in the 
noncommutative deformation of the linear sigma model with 
scalars in the adjoint representation of $U(2)$.  There are several reasons
for this: first, we wish to compare the results for adjoint representation 
scalars with our results from the previous section for fundamental scalars, 
in a tractable case.  Secondly, adjoint matter naturally arises in 
noncommutative world-volume theories on D-branes.  Thirdly, grand unified
theories embedding the standard model commonly rely on adjoints for the
first stage of symmetry breaking.

We write the scalars in the adjoint of $U(2)$ as
\be
\Phi = \phi_{a} T^{a} = \frac{1}{2} 
\left ( \begin{array}{cc} \phi_{4} + \phi_{3} & \sqrt{2} \phi^{*} \\
\sqrt{2} \phi & \phi_{4} - \phi_{3} \end{array} \right )
\label{scalar-rep}
\ee
where $T^{a}$ are the canonical generators of $U(2)$: $T^{a} = 
\sigma^{a}/2$, for $a=1,2,3$ and $T^{4}= I_{2}/2$.
The global $U(2)$ symmetry transformation acts as
\be
\Phi \rightarrow U \Phi U^{\dagger}  
\ee
and as before does not involve the star product because the symmetry is 
global.  For simplicity we impose invariance under $\Phi 
\rightarrow -\Phi$.  The Lagrangian density for the global model we
consider is defined by
\be
{\cal L} = 
\tn{Tr}\left(\partial_{\mu}\Phi*\partial^{\mu}\Phi\right) + 
\mu^{2}\tn{Tr}\left(\Phi*\Phi\right) - 
\lambda_{1}\tn{Tr}\left(\Phi*\Phi*\Phi*\Phi\right) - \lambda_{2} 
\left[\tn{Tr}(\Phi*\Phi)\right]^{2} 
\ee
where we define
\bea
\tn{Tr}( \Phi^{4}_{*} )  &\equiv& \Phi^{i}_{j}\ast \Phi^{j}_{k}\ast 
\Phi^{k}_{l} \ast \Phi^{l}_{i} \nonumber \\
\left[\tn{Tr}(\Phi*\Phi)\right]_{*}^{2}  &\equiv& \Phi^{i}_{j} \ast 
\Phi^{j}_{i} \ast \Phi^{k}_{l} \ast \Phi^{l}_{k}
\eea
and where we discuss the remaining, omitted trace invariants and star
product orderings at the end of this section.

Let us now consider spontaneous symmetry breaking which occurs for 
$\mu^{2}>0$ (we take $\lambda_{i}>0$).  Then 
$\Phi$ acquires a vacuum expectation value, say $\Phi_{0}$, and since it 
is a Hermitian (but not necessarily traceless) matrix, we analyze it by 
diagonalization to the form
\be
\Phi_{0} = \frac{1}{2} 
\left( \begin{array}{cc} a & 0 \\ 0 & b\end{array}\right)
\ee
whence the potential becomes
\be
V(a,b) = -\frac{\mu^{2}}{4} (a^{2} + b^{2}) + 
\frac{\lambda_{1}}{16} (a^{4}+b^{4}) + \frac{\lambda_{2}}{16}(a^{4}+2 
a^{2}b^{2}+b^{4})      
\ee
This is minimized for 
\be
a^{2} = b^{2} = \frac{\mu^{2}}{\frac{\lambda_{1}}{2}+\lambda_{2}} 
\equiv \frac{\mu^{2}}{\lambda}
\label{minima}
\ee
The states corresponding to $a=b$, which are degenerate in energy with 
the states corresponding to $a=-b$, and admitted because we are considering 
$U(2)$ and not simply $SU(2)$, do not reflect spontaneously broken states, 
because $\Phi_{0}$ is then proportional to the identity and so 
manifestly commutes with all of the generators.  Furthermore since 
they correspond to constant shifts in the $U(1)$ component $\phi_{4}$, 
they are forbidden by the discrete symmetry.  On the other hand, the 
states corresponding to $a=-b$ do yield spontaneously broken vacua, 
since they do not commute with the $T^{1}$ and $T^{2}$ generators and 
reflect a vacuum expectation value for the field $\phi_{3}$.

In notation suggestive of the linear sigma model, we expand 
around the vacuum $b=-a<0$ (without any loss of generality), 
defining $\sigma$ and $\pi$ through 
\be
\Phi^{\prime} = \frac{1}{2}  
\left( \begin{array}{cc} \phi_{4} + \sigma & \sqrt{2} \pi^{*}  \\ 
\sqrt{2} \pi & \phi_{4} - \sigma \end{array}\right) \equiv \Phi - 
\Phi_{0} 
\ee
so that $\phi_{3} = \sigma + a$.  Expanding the scalar potential in
terms of these variables yields
\bea
V &=& \frac{1}{2}(2\mu^{2})\sigma^{2} + \frac{1}{2}(\lambda_{1}a^{2})
\phi_{4}^{2} + \frac{\lambda_{1} + \lambda_{2}}{2} \pi^{*}\pi\pi^{*}\pi 
+ \frac{\lambda_{2}}{2}\pi^{*}\pi^{*}\pi\pi  + 
\frac{\lambda_{1} + \lambda_{2}}{2} ( \pi^{*}\pi + \pi 
\pi^{*})\sigma^{2} \nonumber \\ & & - \frac{\lambda_{1}}{2} \pi^{*}\sigma\pi\sigma +
\frac{\lambda_{1} + \lambda_{2}}{2} ( \pi^{*}\pi + 
\pi\pi^{*})\phi_{4}^{2} + \frac{\lambda_{1}}{2} 
\pi^{*}\phi_{4}\pi\phi_{4} + a \lambda (\pi^{*} \pi +\pi \pi^{*} )\sigma 
\nonumber \\ & & + \frac{\lambda}{4} (\sigma^{4}+\phi_{4}^{4}) 
+ \lambda a \sigma^{3} + \frac{\lambda_{1}+\lambda_{2}}{2} \sigma^{2}\phi_{4}^{2} 
+\frac{\lambda_{1}}{4}\sigma\phi_{4}\sigma\phi_{4} + (\lambda + 
\lambda_{1}) a \phi_{4}^{2}\sigma \nonumber \\ & & + \frac{\lambda_{1}}{2}\left[ 
\pi^{*}\phi_{4}\pi\sigma - \pi^{*}\sigma\pi\phi_{4} + a(\pi\pi^{*}\phi_{4}-
\pi^{*}\pi\phi_{4})\right]                    
\label{scalar-potential}
\eea
using $\lambda = \lambda_{1}/2+\lambda_{2}$.  

The symmetrized vertices for this theory are listed in the 
Appendix.  In the following, solid lines denote the $\sigma$, dots denote the 
$\phi_{4}$, and dashes denote the $\pi$.  Excluding the purely noncommutative 
interactions for separate consideration, there are four 1PI graphs 
contributing to the mass renormalization of the complex pion (Goldstone mode) 
in this model:
\begin{fmffile}{global2}
 \bea
   \parbox{120pt}{\begin{fmfgraph*}(120,60)
     \fmfleft{w}
     \fmfright{e}
     \fmf{dashes_arrow,label=$p$}{w,c,e}
     \fmffreeze
     \fmftop{n}
     \fmf{dashes,left,label=$k$}{c,n}
     \fmf{dashes,right}{c,n}
     \fmfdot{c}
     \fmfv{label=$\pi$,label.angle=60}{w}
     \fmfv{label=$\pi^{*}$,label.angle=120}{e}    
    \end{fmfgraph*}} \equiv (a) \hspace{5pt} ; \hspace{5pt}
  \parbox{120pt}{\begin{fmfgraph*}(120,60)
    \fmfleft{w}
    \fmfright{e}
    \fmf{dashes_arrow,label=$p$}{w,c,e}
    \fmffreeze
    \fmftop{n}
    \fmf{plain,left,label=$k$}{c,n}
    \fmf{plain,right}{c,n}
    \fmfdot{c}
    \fmfv{label=$\pi$,label.angle=60}{w}
    \fmfv{label=$\pi^{*}$,label.angle=120}{e}    
   \end{fmfgraph*}} \equiv (b) \hspace{5pt} \nonumber \\
  \parbox{120pt}{\begin{fmfgraph*}(120,60)
    \fmfleft{w}
    \fmfright{e}
    \fmf{dashes_arrow,label=$p$}{w,c,e}
    \fmffreeze
    \fmftop{n}
    \fmf{dots,left,label=$k$}{c,n}
    \fmf{dots,right}{c,n}
    \fmfdot{c}
    \fmfv{label=$\pi$,label.angle=60}{w}
    \fmfv{label=$\pi^{*}$,label.angle=120}{e}    
   \end{fmfgraph*}} \equiv (c) \hspace{5pt} ; \hspace{5pt}
  \parbox{120pt}{\begin{fmfgraph*}(120,60)
    \fmfleft{w}
    \fmfright{e}
    \fmf{dashes_arrow,label=$p$}{w,cw}
    \fmf{dashes,right,tension=0.5,label=$k$}{ce,cw}
    \fmf{plain_arrow,right,tension=0.5,label=$k+p$}{cw,ce}
    \fmf{dashes}{ce,e}
    \fmfdot{cw,ce}
    \fmfv{label=$\pi$,label.angle=60}{w}
    \fmfv{label=$\pi^{*}$,label.angle=120}{e}    
   \end{fmfgraph*}} \equiv (d)
  \eea 
with values given by
\bea
(a) &=& -2 i (\lambda_{1} + \lambda_{2}) \int \frac{d^{4}k}{(2\pi)^{4}} 
    \frac{i}{k^{2}} - 2 i \lambda_{2} \int \frac{d^{4}k}{(2\pi)^{4}} 
    \frac{ i \cos^{2}(\frac{k\times p}{2})}{k^{2}} \nonumber \\     
    &=& (2\lambda_{1}  + 3\lambda_{2})I(0) + 
    \lambda_{2}I_{\theta,p}(0) \nonumber \\
(b) &=& (\lambda_{1}  + \lambda_{2})I(2\mu^{2}) - 
    \frac{\lambda_{1}}{2}I_{\theta,p}(2\mu^{2} ) \nonumber \\
(c) &=& (\lambda_{1} + \lambda_{2}) I(\lambda_{1}a^{2}) + 
    \frac{\lambda_{1}}{2} I_{\theta,p}(\lambda_{1}a^{2}) \nonumber \\
(d) &=& (-2 i \lambda a)^{2} \int \frac{d^{4}k}{(2\pi)^{4}} 
  \frac{i}{k^{2}} \frac{i}{(p+k)^{2} - 2\mu^{2}} \cos^{2}
  (\frac{k\times p}{2})  \nonumber \\
   &=& 4 \lambda^{2} a^{2} \int \frac{d^{4}k}{(2\pi)^{4}}
   \frac{1}{2\mu^{2}} \left[ \frac{1}{k^{2}-2\mu^{2}}  - \frac{1}{k^{2}} 
   \right] \cos^{2} (\frac{k\times p}{2}) + C^{\mu}(p) p_{\mu}  \nonumber \\
   &=& \lambda \left[ I(2\mu^{2}) - I(0) \right] + \lambda \left[ 
   I_{\theta,p}(2\mu^{2})  - I_{\theta,p}(0) \right] + C^{\mu}(p) p_{\mu}    
\eea
where $C^{\mu}$ is finite for all $p$.

The one-point tadpole contributions are given by
\bea
   \parbox{100pt}{\begin{fmfgraph*}(100,60)
     \fmfleft{w}
     \fmfright{e}
     \fmf{dashes_arrow,label=$p$}{w,c,e}
     \fmffreeze
     \fmftop{n}
     \fmf{plain,tension=2.5}{c,cn}
     \fmf{dashes,right}{cn,n}
     \fmf{dashes,left}{cn,n}
     \fmfdot{c}
     \fmfdot{cn}
     \fmfv{label=$\pi$,label.angle=60}{w}
     \fmfv{label=$\pi^{*}$,label.angle=120}{e}    
    \end{fmfgraph*}} \equiv (e) ;
  \parbox{100pt}{\begin{fmfgraph*}(100,60)
     \fmfleft{w}
     \fmfright{e}
     \fmf{dashes_arrow,label=$p$}{w,c,e}
     \fmffreeze
     \fmftop{n}
     \fmf{plain,tension=2.5}{c,cn}
     \fmf{plain,right}{cn,n}
     \fmf{plain,left}{cn,n}
     \fmfdot{c}
     \fmfdot{cn}
     \fmfv{label=$\pi$,label.angle=60}{w}
     \fmfv{label=$\pi^{*}$,label.angle=120}{e}    
    \end{fmfgraph*}} \equiv (f) ; 
   \parbox{100pt}{\begin{fmfgraph*}(100,60)
     \fmfleft{w}
     \fmfright{e}
     \fmf{dashes_arrow,label=$p$}{w,c,e}
     \fmffreeze
     \fmftop{n}
     \fmf{plain,tension=2.5}{c,cn}
     \fmf{dots,right}{cn,n}
     \fmf{dots,left}{cn,n}
     \fmfdot{c}
     \fmfdot{cn}
     \fmfv{label=$\pi$,label.angle=60}{w}
     \fmfv{label=$\pi^{*}$,label.angle=120}{e}    
    \end{fmfgraph*}} \equiv (g) 
\eea    
which are respectively given by
\bea
(e) &=& (-2 i \lambda a)^{2} \frac{i}{-2\mu^{2}} \hspace{5pt} i I(0) 
= -2 \lambda I(0) \nonumber \\
(f) &=& (-2 i \lambda a) \frac{i}{-2\mu^{2}} (-6i\lambda a) \frac{i}{2}
    I(2\mu^{2}) = -3 \lambda I(2\mu^{2}) \nonumber \\
(g) &=& (-2 i \lambda a) \frac{i}{-2\mu^{2}} (-2 i (\lambda + \lambda_{1})
      a ) \frac{i}{2} I(\lambda_{1}a^{2}) \nonumber \\
    &=& - (\lambda + \lambda_{1}) I(\lambda_{1}a^{2})
\eea

The sum of these seven graphs is given by 
\be
\sum = \frac{\lambda_{1}}{2} \left[ I(0) - I_{\theta,p}(0) \right] - 
\lambda_{2} \left[ I(2\mu^{2}) - I_{\theta,p}(2\mu^{2})\right] -
\frac{\lambda_{1}}{2} \left[ I(\lambda_{1}a^{2}) - 
I_{\theta,p}(\lambda_{1}a^{2}) \right] + C^{\mu}(p) p_{\mu}
\label{partial-sum}
\ee
In the commutative limit $\theta \rightarrow 0$, this degenerates to the 
finite term $C^{\mu}(p) p_{\mu}$ (which itself vanishes as $p\rightarrow 0$), 
so the mass counterterm vanishes and this is a demonstration of 
Goldstone's theorem for this model.  However for nonzero $\theta$, 
the $I(m^{2})$ terms are divergent and require regularization, 
say by an ultraviolet cutoff $\Lambda$.  But there is 
no counterterm freedom to cancel the $\Lambda$ dependence, so for
nonzero $p$ and nonzero $\theta$ we cannot take the continuum limit; 
that is, UV ($\Lambda\rightarrow \infty$) and IR ($p\rightarrow 0$) limits
do not commute.

However, we have (intentionally) neglected a purely noncommutative graph due 
to the last interaction in (\ref{scalar-potential}).  The purely 
noncommutative interaction generated by $(\pi \pi^{*} - \pi^{*} \pi) 
\phi_{4}$ yields a graphical contribution given by

\bea
  \parbox{100pt}{\begin{fmfgraph*}(100,60)
    \fmfleft{w}
    \fmfright{e}
    \fmf{dashes_arrow,label=$p$}{w,cw}
    \fmf{dashes_arrow,right,tension=0.5,label=$k$}{ce,cw}
    \fmf{dots,right,tension=0.5,label=$k+p$}{cw,ce}
    \fmf{dashes}{ce,e}
    \fmfdot{cw,ce}
    \fmfv{label=$\pi$,label.angle=60}{w}
    \fmfv{label=$\pi^{*}$,label.angle=120}{e}    
  \end{fmfgraph*}}
  &=& (-\lambda_{1} a)^{2} i^{2} \int \frac{d^{4}k}{(2\pi)^{4}} 
  \frac{\sin(\frac{p\times k}{2}) \sin(\frac{-k\times 
  -p}{2})}{k^{2} [ (p+k)^{2} - \lambda_{1} a^{2}]} \nonumber \\
  &=& \lambda_{1}^{2} a^{2} \int \frac{d^{4}k}{(2\pi)^{4}} 
  \frac{\sin^{2}(\frac{k\times p}{2})}{\lambda_{1}a^{2}} \left[
  \frac{1}{k^{2}-\lambda_{1}a^{2}} - \frac{1}{k^{2}} \right] + 
  D_{\theta}^{\mu}(p) p_{\mu} 
  \nonumber \\
  &=& \frac{\lambda_{1}}{2} \int \frac{d^{4}k}{(2\pi)^{4}} \left[
   1- \cos(k\times p) \right] \left[
  \frac{1}{k^{2}-\lambda_{1}a^{2}} - \frac{1}{k^{2}} \right] + 
  D_{\theta}^{\mu}(p) p_{\mu} \nonumber \\
  &=& \frac{\lambda_{1}}{2} \left[ I(\lambda_{1}a^{2}) - 
  I_{\theta,p}(\lambda_{1}a^{2}) \right] - \frac{\lambda_{1}}{2} 
  \left[ I(0) - I_{\theta,p}(0) \right] + D_{\theta}^{\mu}(p) p_{\mu}          
\eea
\end{fmffile}
\noi
where again $D^{\mu}_{\theta}$ is finite for all $p$, and vanishes also
in the limit $\theta\rightarrow 0$. Rather unexpectedly, this graph, which 
manifestly vanishes in the commutative limit, and involves the 
$U(1)$ component of the matter field, cancels the $\lambda_{1}$ pieces in 
(\ref{partial-sum}), leaving behind a residual divergence (for nonzero $p$) 
that depends only on the coupling to the $\tn{Tr}(\Phi^{2})^{2}$ term in the 
potential.   

However, in the corresponding gauge theory, the term
\be 
\tn{Tr}(\Phi^{2}_{\ast})\ast \tn{Tr} (\Phi^{2}_{\ast})
\label{two-trace}
\ee 
is {\it not} gauge invariant even under the spacetime integral.  
In fact no term involving the product of more than one trace in the 
adjoint representation is gauge invariant (even under $\int d^{D}x$) in 
noncommutative theories for $N>1$.  To see this write (\ref{two-trace}) in 
terms of its internal indices (first choosing the canonical ordering with
respect to the star product) and gauge transform:
\bea
[\tn{Tr}(\Phi^{2}_{\ast})]^{2}_{*} &=& \Phi^{i}_{j}\ast \Phi^{j}_{i}\ast
\Phi^{k}_{l}\ast \Phi^{l}_{k} \nonumber \\
&\rightarrow& ( U^{i}_{i1}\ast \Phi^{i1}_{j1} \ast U^{\dagger 
j1}_{j} \ast U^{j}_{j2}\ast \Phi^{j2}_{i2} \ast U^{\dagger i2}_{i}) \ast
( U^{k}_{k1}\ast \Phi^{k1}_{l1} \ast U^{\dagger l1}_{l} \ast
U^{l}_{l2}\ast \Phi^{l2}_{k2} \ast U^{\dagger k2}_{k} ) \nonumber \\
&=& U^{i}_{i1} \ast\Phi^{i1}_{j1}\ast\Phi^{j1}_{i2} \ast U^{\dagger 
i2}_{i} \ast U^{k}_{k1} \ast\Phi^{k1}_{l1}\ast\Phi^{l1}_{k2} \ast U^{\dagger 
k2}_{k}        
\eea
The presence of the star product does not allow us to use
$U^{i}_{j}\ast U^{\dagger j}_{k}= \delta^{i}_{k}$ on the remaining {\it local} 
$U$ and $U^{\dagger}$ factors which are separated by two factors of 
$\Phi$, even if we use the cyclicity property of the star product under 
the spacetime integral.  This is to be contrasted with a single internal 
index trace term (with canonical internal index ordering), and the 
commutative limit where the ordering of components is immaterial.

It is clear this argument applies both to the other internal index 
ordering $\Phi^{i}_{j}\ast \Phi^{k}_{l}\ast \Phi^{j}_{i}\ast 
\Phi^{l}_{k}$    
(whose gauge transformation does not allow the use of $U\ast U^{\dagger}=I$ 
anywhere), and to any product of (internal index) traces in the adjoint 
representation.  Thus if we forbid $[\tn{Tr}(\Phi^{2})]^{2}$ from the 
scalar potential, by regarding the global theory as the limit of a 
gauge theory, we have no remaining violation of Goldstone's theorem for this 
model.  Incidentally, this argument also forbids the other terms 
still allowed by the imposition of the discrete symmetry that 
we neglected when we wrote the scalar potential for this theory; namely
\be
\tn{Tr}(\Phi)\ast \tn{Tr}(\Phi^{3}) \hspace{5pt}, \hspace{5pt}
\tn{Tr}(\Phi)\ast \tn{Tr}(\Phi) \hspace{5pt},\hspace{5pt}
[\tn{Tr}(\Phi)]^{4}_{\ast}\hspace{5pt},\hspace{5pt} 
\tn{Tr}(\Phi^{2}_{\ast})\ast[\tn{Tr}(\Phi)]^{2}_{\ast}
\ee
as well as other star product orderings of the $\tn{Tr}(\Phi^{4})$ term.

An immediate consequence of the preceding argument is that for $U(N)$ gauge
theories with adjoint scalar matter, the symmetry breaking pattern is
restricted to only one of the two possible patterns that would be allowed
by the commutative limit of the theory.  Specifically, because noncommutative
gauge invariance forbids $\tn{Tr}(\Phi^{2}_{*})\ast \tn{Tr}(\Phi^{2}_{*})$,
vacuum stability now requires $\lambda_{1}>0$, and thus allows only the 
breaking pattern $U(N)\rightarrow U(n_{1})\times U(N-n_{1})$ (with
$n_{1}= N/2$, $N$ even; or $n_{1}= (N+1)/2$, $N$ odd)\cite{l}, and forbids
$U(N)\rightarrow U(N-1)$\cite{l}.  

This argument has another consequence for noncommutative theories in 
general.  As van Raamsdonk and Seiberg \cite{rs} demonstrated in 
considering scalar theories with scalars represented by $N\times N$
matrices, all infrared divergences of the type found in \cite{mrs}
are proportional at one-loop to
\be
\tn{Tr}({\cal O}_{1}) \tn{Tr}({\cal O}_{2})   
\ee
where ${\cal O}_{i}$ are operators built out of $\Phi$.  Furthermore we have
seen above that an operator of this form ($\tn{Tr}(\Phi_{*}^{2})\ast
\tn{Tr}(\Phi_{*}^{2})$) appearing in the scalar potential, would induce
violations of Goldstone's theorem by renormalization effects. 
However the preceding argument indicates that these are precisely the form of 
operators that are not gauge invariant in an adjoint representation gauge 
theory.  So if we regard these theories as 
embedded in a corresponding gauge theory where we must forbid such terms, 
then we would expect that infrared divergences for $N>1$\footnote{For the 
$N=1$ case considered in \cite{mrs}, corresponding to a single 
scalar, the above argument fails, since the index structure 
becomes degenerate.} of the form observed in \cite{rs}, no longer appear.

\section{NC O(4) Sigma Model: Adjoint Representation}

In this section we repeat the analysis of the previous section for the 
noncommutative $O(4)$ sigma model in the adjoint representation; again this
will allow us to study, in a simple context, scalar representation 
(in)dependence of our results on Goldstone renormalization, this time in the 
context of orthogonal symmetry groups.

We consider the classical symmetry breaking $O(4)\rightarrow U(2)$.
Now $\Phi$ is a real antisymmetric matrix, whence the vacuum state
$\Phi_{0}$ can be put in standard form
\be
\Phi_{0} = \frac{a}{2} 
\left( \begin{array}{cccc} 0 & 1 & 0 & 0 \\ -1 & 0 & 0 & 0 \\
0 & 0 & 0 & 1 \\ 0 & 0 & -1 & 0 \end{array}\right)
\label{vac}
\ee

The scalar potential is given by 
\be
V(\Phi) = \frac{\mu^2}{2} \tn{Tr}(\Phi^{2}_{*}) + \frac{\lambda_{1}}{4} 
[\tn{Tr}(\Phi^{2}_{*})]^{2}_{*} + \frac{\lambda_{2}}{4} \tn{Tr}(\Phi^{4}_{*}) 
\ee
where we note that the sign of the quadratic term is opposite that of the 
$U(2)$ model of the previous section because of the antisymmetry (as opposed
to Hermicity) of $\Phi$, and where we have normalized differently for 
later convenience (we now assume the canonical internal index ordering 
with respect to the star product as per the conclusions of the previous
section).  Thus the minimization of the potential with respect to the
vacuum (\ref{vac}), yields
\be
V(\Phi_{0}) = - \frac{\mu^{2}}{2} a^{2} + \frac{\lambda_{1} a^{4}}{4} +
\frac{\lambda_{2} a^{4}}{16} \hspace{5pt} \rightarrow \hspace{5pt}
a^{2} = \frac{4 \mu^{2}}{4 \lambda_{1} + \lambda_{2}} = \frac{\mu^{2}}{
\lambda_{1} + \frac{\lambda_{2}}{4}}
\ee
The suitable parametrization of $\Phi$ relevant to a discussion of spontaneous
symmetry breaking is given by
\be
\frac{1}{2}
\left( \begin{array}{cccc} 0 & (\sigma + a) + \psi & \alpha + \pi_{1} & 
\beta + \pi_{2} \\ -((\sigma + a) + \psi) & 0 & \pi_{2} - \beta & \alpha
- \pi_{1} \\ -(\alpha+ \pi_{1}) & \beta - \pi_{2} & 0 & (\sigma+a)-\psi \\ 
-(\beta + \pi_{2}) & \pi_{1}-\alpha & \psi-(\sigma+a) & 0 \end{array}\right)
\ee
where the $\sigma$ is the field acquiring the VEV, and $\pi_{1},\pi_{2}$ are 
the two Goldstone modes.  Focussing now on the one-loop mass renormalization 
of one of the $\pi$'s, say
$\pi_{1}$, the expansion of the potential reads
\bea
V &=& \frac{1}{2} (2\mu^{2}) \sigma^{2} + \frac{1}{2}(\lambda_{2}a^{2}/2)
\left[\psi^{2} + \alpha^{2} + \beta^{2}\right] + \frac{1}{4}\left(
\lambda_{1}+\frac{\lambda_{2}}{4}\right) \pi_{1}^{4} + 
\left(\frac{\lambda_{1}}{2}+\frac{\lambda_{2}}{4}\right)\pi_{1}^{2}
\left[ \pi_{2}^{2} + \right. \nonumber \\ & & \left. \alpha^{2} + \beta^{2}
+ \psi^{2} + \sigma^{2}\right] + \frac{\lambda_{2}}{8} \bigg[ - \pi_{1} 
\pi_{2} \pi_{1} \pi_{2} + \pi_{1} \alpha \pi_{1} \alpha + \pi_{1} 
\beta \pi_{1} \beta + \pi_{1}\psi \pi_{1} \psi - \pi_{1} \sigma \pi_{1} 
\sigma \bigg] \nonumber \\
& & + \left( \lambda_{1} + \frac{\lambda_{2}}{4}\right) a \sigma \left[ 
\sigma^{2} + \pi_{1}^{2} + \pi_{2}^{2} \right] +  \left( \lambda_{1}
+ \frac{3 \lambda_{2}}{4}\right) a \sigma \left[ \alpha^{2} + \beta^{2}
+ \psi^{2} \right] + \ldots
\eea
where the ellipsis represents (four-field) terms that do not contribute to
the one-loop mass renormalization of $\pi_{1}$.  The Feynman rules for these
vertices are in the appendix.  Now that we have six distinct fields, we simply
use dotted lines to denote the $\pi$'s, and use solid lines for the other four
fields, and instead explicitly label the lines.

The 1PI graphs contributing are
\begin{fmffile}{global3}
 \bea
& &   \parbox{100pt}{\begin{fmfgraph*}(100,60)
     \fmfleft{w}
     \fmfright{e}
     \fmf{dashes_arrow,label=$p$}{w,c,e}
     \fmffreeze
     \fmftop{n}
     \fmf{dashes,left,label=$\pi_{1}$}{c,n}
     \fmf{dashes,right}{c,n}
     \fmfdot{c}
     \fmfv{label=$\pi_{1}$,label.angle=60}{w}
     \fmfv{label=$\pi_{1}$,label.angle=120}{e}    
    \end{fmfgraph*}} \equiv (a) \hspace{5pt} ; \hspace{5pt}
   \parbox{100pt}{\begin{fmfgraph*}(100,60)
     \fmfleft{w}
     \fmfright{e}
     \fmf{dashes_arrow,label=$p$}{w,c,e}
     \fmffreeze
     \fmftop{n}
     \fmf{dashes,left,label=$\pi_{2}$}{c,n}
     \fmf{dashes,right}{c,n}
     \fmfdot{c}
     \fmfv{label=$\pi_{1}$,label.angle=60}{w}
     \fmfv{label=$\pi_{1}$,label.angle=120}{e}    
    \end{fmfgraph*}} \equiv (b) \hspace{5pt} ; \hspace{5pt}
  \parbox{100pt}{\begin{fmfgraph*}(100,60)
    \fmfleft{w}
    \fmfright{e}
    \fmf{dashes_arrow,label=$p$}{w,c,e}
    \fmffreeze
    \fmftop{n}
    \fmf{plain,left,label=$\alpha(\beta)(\psi)$}{c,n}
    \fmf{plain,right}{c,n}
    \fmfdot{c}
    \fmfv{label=$\pi_{1}$,label.angle=25}{w}
    \fmfv{label=$\pi_{1}$,label.angle=120}{e}    
   \end{fmfgraph*}} \equiv (c) \hspace{5pt} \nonumber \\
& &  \parbox{100pt}{\begin{fmfgraph*}(100,60)
    \fmfleft{w}
    \fmfright{e}
    \fmf{dashes_arrow,label=$p$}{w,c,e}
    \fmffreeze
    \fmftop{n}
    \fmf{plain,left,label=$\sigma$}{c,n}
    \fmf{plain,right}{c,n}
    \fmfdot{c}
    \fmfv{label=$\pi_{1}$,label.angle=60}{w}
    \fmfv{label=$\pi_{1}$,label.angle=120}{e}    
   \end{fmfgraph*}} \equiv (d) \hspace{5pt} ; \hspace{5pt}
  \parbox{100pt}{\begin{fmfgraph*}(100,60)
    \fmfleft{w}
    \fmfright{e}
    \fmf{dashes_arrow,label=$p$}{w,cw}
    \fmf{dashes,right,tension=0.5,label=$\pi_{1}$}{ce,cw}
    \fmf{plain_arrow,right,tension=0.5,label=$\sigma$}{cw,ce}
    \fmf{dashes}{ce,e}
    \fmfdot{cw,ce}
    \fmfv{label=$\pi_{1}$,label.angle=60}{w}
    \fmfv{label=$\pi_{1}$,label.angle=120}{e}    
   \end{fmfgraph*}} \equiv (e)
  \eea 
and are given respectively by
\bea
(a) &=& -2 i \left(\lambda_{1} + \frac{\lambda_{2}}{4}\right)
\frac{i}{2} \int \frac{d^{4}k}{(2\pi)^{4}} \frac{1 + \cos^{2}(\frac{k\times
p}{2})}{k^{2}} = (\lambda_{1} + \frac{\lambda_{2}}{4}) \left[ 2 I(0) + 
I_{\theta,p}(0) \right] \nonumber \\
(b) &=& \left( \lambda_{1} + \frac{\lambda_{2}}{2}\right) I(0) - \frac{
\lambda_{2}}{4} I_{\theta,p}(0) \nonumber \\
(c) &=& 3 \left[ \left(\lambda_{1}+\frac{\lambda_{2}}{2}\right) 
I(\lambda_{2} a^{2}/2) + \frac{\lambda_{2}}{4} I_{\theta,p}(\lambda_{2}
a^{2}/2) \right] \nonumber \\
(d) &=& \left( \lambda_{1} + \frac{\lambda_{2}}{2}\right) I(2\mu^{2}) 
- \frac{\lambda_{2}}{4} I_{\theta,p}(2\mu^{2}) \nonumber \\
(e) &=& \left[ -2i \left( \lambda_{1} + \frac{\lambda_{2}}{4}\right) a
\right]^{2} i^{2} \int \frac{d^{4}k}{(2\pi)^{4}} \frac{\cos^{2}(\frac{
p\times k}{2})}{k^{2} [(p+k)^{2} - 2 \mu^{2}]} \nonumber \\
&=& \frac{2  \left( \lambda_{1} + \frac{\lambda_{2}}{4}\right)^{2} a^{2}}
{2\mu^{2}} \int \frac{d^{4}k}{(2\pi)^{4}} \left[ 1 + \cos(p\times k)\right]
\left[ \frac{1}{k^{2}-2\mu^{2}} - \frac{1}{k^{2}}\right] + D^{\mu} p_{\mu}
\nonumber \\
&=& \left( \lambda_{1} + \frac{\lambda_{2}}{4}\right) \left[ I(2\mu^{2})
- I(0)\right]  + \left( \lambda_{1} + \frac{\lambda_{2}}{4}\right) 
\left[ I_{\theta,p}(2\mu^{2})- I_{\theta,p}(0)\right] + D^{\mu} p_{\mu} 
\eea
where $D^{\mu}$ is finite for all p, and where the factor of three in the 
third graph originates from having three species of particle with the 
same contribution to this calculation.

The one-point tadpoles contributions are
\bea
   \parbox{100pt}{\begin{fmfgraph*}(100,60)
     \fmfleft{w}
     \fmfright{e}
     \fmf{dashes_arrow,label=$p$}{w,c,e}
     \fmffreeze
     \fmftop{n}
     \fmf{plain,tension=2.5,label=$\sigma$}{c,cn}
     \fmf{dashes,right}{cn,n}
     \fmf{dashes,left,label=$\pi_{1(2)}$}{cn,n}
     \fmfdot{c}
     \fmfdot{cn}
     \fmfv{label=$\pi_{1}$,label.angle=60}{w}
     \fmfv{label=$\pi_{1}$,label.angle=120}{e}    
    \end{fmfgraph*}} \equiv (f) ;
  \parbox{100pt}{\begin{fmfgraph*}(100,60)
     \fmfleft{w}
     \fmfright{e}
     \fmf{dashes_arrow,label=$p$}{w,c,e}
     \fmffreeze
     \fmftop{n}
     \fmf{plain,tension=2.5,label=$\sigma$}{c,cn}
     \fmf{plain,right}{cn,n}
     \fmf{plain,left,label=$\sigma$}{cn,n}
     \fmfdot{c}
     \fmfdot{cn}
     \fmfv{label=$\pi_{1}$,label.angle=60}{w}
     \fmfv{label=$\pi_{2}$,label.angle=120}{e}    
    \end{fmfgraph*}} \equiv (g) ; 
   \parbox{100pt}{\begin{fmfgraph*}(100,60)
     \fmfleft{w}
     \fmfright{e}
     \fmf{dashes_arrow,label=$p$}{w,c,e}
     \fmffreeze
     \fmftop{n}
     \fmf{plain,tension=2.5,label=$\sigma$}{c,cn}
     \fmf{plain,right}{cn,n}
     \fmf{plain,left,label=$\alpha(\beta)(\psi)$}{cn,n}
     \fmfdot{c}
     \fmfdot{cn}
     \fmfv{label=$\pi_{1}$,label.angle=60}{w}
     \fmfv{label=$\pi_{1}$,label.angle=120}{e}    
    \end{fmfgraph*}} \equiv (h)
\eea
\end{fmffile}
with values given by
\bea
(f) &=& 2 \times -2 i \left( \lambda_{1} + \frac{\lambda_{2}}{4}\right)
a \frac{i}{-2\mu^{2}} (-2 i) \left( \lambda_{1} + \frac{\lambda_{2}}{4}\right)
a \frac{i}{2} I(0) = -2 \left( \lambda_{1} + \frac{\lambda_{2}}{4}\right) I(0)
\nonumber \\
(g) &=& -2 i \left( \lambda_{1} + \frac{\lambda_{2}}{4}\right)
a \frac{i}{-2\mu^{2}} (-6 i) \left( \lambda_{1} + \frac{\lambda_{2}}{4}\right)
a \frac{i}{2} I(2\mu^{2}) = -3 \left( \lambda_{1} + \frac{\lambda_{2}}{4}\right) I(2\mu^{2}) \nonumber \\
(h) &=& 3 \times -2 i \left( \lambda_{1} + \frac{\lambda_{2}}{4}\right)
a \frac{i}{-2\mu^{2}} (-2 i) \left( \lambda_{1} + \frac{3 \lambda_{2}}{4}
\right)
a \frac{i}{2} I(\frac{\lambda_{2}a^{2}}{2}) = -3 \left( \lambda_{1} + 
\frac{3 \lambda_{2}}{4}\right) I(\frac{\lambda_{2}a^{2}}{2}) \nonumber \\
\eea
where the overall factors of two and three in the first and third 
graphs respectively again come from the multiplicity of particle species with the same contribution.

Thus the total one-loop contribution to the mass renormalization of the 
$\pi_{1}$ (or $\pi_{2}$) in this model is
\be
\sum = \frac{\lambda_{2}}{4} \left[ I(0) - I_{\theta,p}(0) \right]
- \frac{3\lambda_{2}}{4} \left[ I(\frac{\lambda_{2}a^{2}}{2}) - I_{\theta,p}
(\frac{\lambda_{2}a^{2}}{2}) \right] - \lambda_{1} \left[ I(2\mu^{2}) - I_{\theta,p}
(2\mu^{2}) \right] + D^{\mu} p_{\mu}
\ee
Unlike the $U(2)$ adjoint representation model, there is no purely
noncommutative graph that saves us for either quartic invariant, and
so again we cannot take the continuum limit ($\Lambda_{UV}\rightarrow
\infty$) and Goldstone's theorem fails for this model.


\section{Discussion}

To summarize: in noncommutative field theory, $U(N)$ ($N>1$) linear sigma
models with complex scalars in the fundamental representation, do not
have $O(2N)$ global invariance due to noncommutative commutator interactions
between the real components, which vanish in the commutative limit.
As a result of these commutator interactions, noncommutative linear $U(N)$
sigma models with fundamental matter can be continuum renormalized while
preserving Nambu-Goldstone symmetry realization, at least at one-loop.  This
contrasts with our previous results \cite{us}, where we demonstrated
that for noncommutative linear $O(N)$ sigma models with fundamental matter,
continuum renormalization is inconsistent with Nambu-Goldstone symmetry
realization already at one loop (except for the degenerate Abelian case
$O(2)\equiv U(1)$.

To investigate possible scalar representation dependence of these contrasting
results, we have considered linear sigma models with adjoint matter.  For the
adjoint $U(2)$ linear sigma model, we again find that Nambu-Goldstone
symmetry realization survives at one-loop, provided we drop interaction
terms (and star product orderings) which would be inconsistent with the
gauging of the symmetry; noncommutative restrictions on the allowed operators
in a $U(N)$ gauge theory Lagrangian also act to restrict the allowed symmetry
breaking patterns.  For the adjoint $O(4)$ linear sigma model, we find 
violations of Nambu-Goldstone symmetry realization at one-loop order, as in
the fundamental $O(N)$ models.  These results suggest that the difference in
behaviour is determined by the symmetry group, as opposed to the scalar 
representation thereof.

Our results for the noncommutative linear $U(N)$ sigma models now open the 
possibility of building models of the elementary particles
and their interactions based on noncommutative non-Abelian theories with
spontaneous symmetry breaking.  Clearly, to make particle physics models,
it is necessary that the spontaneous symmetry breaking be consistent with
the renormalization not just of the global limits of these theories, but also
with their gaugings; we see two reasons to be sanguine on this point, at least
at one-loop.  First, the gauging of the $U(1)$($=O(2)$)) model\cite{p} is
consistent with spontaneous symmetry breaking for precisely the star orderings
uniquely picked out\footnote{to make the anomalous effects vanish, which
can happen only in the Abelian $O(2)$ case.} by our\cite{us} calculation 
of the Goldstone violating effects in the general noncommutative $O(N)$ 
fundamental linear sigma model.  Second, in our treatment of the non-Abelian
$U(2)$ model with adjoint scalars, violations of Nambu-Goldstone symmetry
realization vanish when one restricts to the subset of couplings which would
be allowed, were the symmetry to be gauged; so the limited evidence suggests
that global theories may be a good guide to the behaviour of the local 
theories, much as in the case of commutative field theories\cite{ssb1},
\cite{ssb2}.

However, to go from models to actual theories would require demonstration of
all-order consistency of continuum renormalization of noncommutative theories
with spontaneous symmetry breaking.  While failure of Nambu-Goldstone 
symmetry breaking can be demonstrated at one-loop, demonstrating consistency
requires an all order analysis; this remains a major open issue in this
field.

\section*{Acknowledgements}

We would like to thank Andrzej Czarnecki for a useful conversation.  This
research is supported in part by the Natural Sciences and Engineering Research
Council of Canada.

\appendix

\section{Appendix}

\subsection{Scalar Potential Feynman Rules, $U(N)$ Fundamental}

All momenta flow into the vertices.

\begin{fmffile}{vertices1}
  \bea
    \parbox{70pt}{\begin{fmfgraph*}(50,50)
     \fmfsurroundn{v}{8}
     \fmfdot{c}
     \fmf{dashes}{v2,c}
     \fmf{dashes}{v4,c}
     \fmf{dashes}{v6,c}
     \fmf{dashes}{v8,c} 
     \fmfv{label=$\pi_{j} \hspace{5pt} p_{2}$}{v2}
     \fmfv{label=$\pi^{*}_{i} \hspace{5pt} p_{1}$}{v4}
     \fmfv{label=$\pi^{*}_{k}  \hspace{5pt} p_{3}$}{v6}
     \fmfv{label=$\pi_{l} \hspace{5pt} p_{4}$}{v8}
  \end{fmfgraph*}} & = & - 2 i \lambda \left[ \delta^{ij}\delta^{kl}
      \tn{e}^{-\frac{i}{2}(p_{1}\times p_{2} + p_{3}\times p_{4})} 
      + \delta^{il}\delta^{jk} \tn{e}^{+\frac{i}{2}(p_{1}\times 
      p_{2} + p_{3}\times p_{4})} \right]  \nonumber \\ \\ \nonumber \\
  \parbox{70pt}{\begin{fmfgraph*}(50,50)
     \fmfsurroundn{v}{8}
     \fmfdot{c}
     \fmf{plain}{v2,c}
     \fmf{plain}{v4,c} 
     \fmf{plain}{v6,c}
     \fmf{plain}{v8,c}
     \fmfv{label=$\sigma(\pi_{0}) \hspace{5pt} p_{2}$}{v2}
     \fmfv{label=$\sigma(\pi_{0}) \hspace{5pt} p_{1}$}{v4}
     \fmfv{label=$\sigma(\pi_{0}) \hspace{5pt} p_{3}$}{v6}
     \fmfv{label=$\sigma(\pi_{0}) \hspace{5pt} p_{4}$}{v8}
  \end{fmfgraph*}} & = & - 2 i \lambda \left[ \cos(\frac{p_{1} \times p_{2}}{2})
    \cos(\frac{p_{3} \times p_{4}}{2}) + \cos(\frac{p_{1} \times p_{3}}{2})
    \cos(\frac{p_{2} \times p_{4}}{2}) \right. \nonumber \\ & & \left. + \cos(\frac{p_{1} \times 
    p_{4}}{2}) \cos(\frac{p_{2} \times p_{3}}{2}) \right] \nonumber \\ \\ \nonumber \\
  \parbox{70pt}{\begin{fmfgraph*}(50,50)
     \fmfsurroundn{v}{8}
     \fmfdot{c}
     \fmf{plain}{v2,c}
     \fmf{plain}{v4,c} 
     \fmf{dots}{v6,c}
     \fmf{dots}{v8,c}
     \fmfv{label=$\sigma \hspace{5pt} p_{2}$}{v2}
     \fmfv{label=$\sigma \hspace{5pt} p_{1}$}{v4}
     \fmfv{label=$\pi_{0} \hspace{5pt} p_{3}$}{v6}
     \fmfv{label=$\pi_{0} \hspace{5pt} p_{4}$}{v8}
  \end{fmfgraph*}} & = & - 2 i \lambda [ 2
    \cos(\frac{p_{1} \times p_{2}}{2}) \cos(\frac{p_{3} \times 
    p_{4}}{2}) - \cos(\frac{p_{1} \times p_{3}}{2} + 
    \frac{p_{2} \times p_{4}}{2}) ] \nonumber \\ \\ \nonumber \\
  \parbox{70pt}{\begin{fmfgraph*}(50,50)
     \fmfsurroundn{v}{8}
     \fmfdot{c}
     \fmf{dashes}{v2,c}
     \fmf{dashes}{v4,c} 
     \fmf{plain}{v6,c}
     \fmf{plain}{v8,c}
     \fmfv{label=$\pi_{j} \hspace{5pt} p_{2}$}{v2}
     \fmfv{label=$\pi^{*}_{i} \hspace{5pt} p_{1}$}{v4}
     \fmfv{label=$\sigma(\pi_{0}) \hspace{5pt} p_{3}$}{v6}
     \fmfv{label=$\sigma(\pi_{0}) \hspace{5pt} p_{4}$}{v8}
  \end{fmfgraph*}} & = & - 2 i \lambda \delta^{ij} \tn{e}^{-\frac{i}{2}(p_{1} 
      \times p_{2})} \cos(\frac{p_{3}\times p_{4}}{2})    
    \nonumber \\ \\ \nonumber \\
  \parbox{70pt}{\begin{fmfgraph*}(50,50)
     \fmfsurroundn{v}{8}
     \fmfdot{c}
     \fmf{dashes}{v2,c}
     \fmf{dashes}{v4,c} 
     \fmf{plain}{v6,c}
     \fmf{dots}{v8,c}
     \fmfv{label=$\pi_{j} \hspace{5pt} p_{2}$}{v2}
     \fmfv{label=$\pi^{*}_{i} \hspace{5pt} p_{1}$}{v4}
     \fmfv{label=$\sigma \hspace{5pt} p_{3}$}{v6}
     \fmfv{label=$\pi_{0} \hspace{5pt} p_{4}$}{v8}
  \end{fmfgraph*}} & = & - 2 i \lambda \delta^{ij} \tn{e}^{-\frac{i}{2}(p_{1} 
      \times p_{2})} \sin(\frac{p_{3}\times p_{4}}{2}) \\ \nonumber \\ \nonumber \\   
  \parbox{70pt}{\begin{fmfgraph*}(50,50)
     \fmfsurround{v1,v2,v3}
     \fmfdot{c}
     \fmf{plain}{v1,c}
     \fmf{plain}{v2,c}
     \fmf{plain}{v3,c}
     \fmfv{label=$\sigma \hspace{5pt} p_{3}$,label.angle=60}{v1}
     \fmfv{label=$\sigma \hspace{5pt} p_{1}$,label.angle=180}{v2}
     \fmfv{label=$\sigma \hspace{5pt} p_{2}$}{v3}
  \end{fmfgraph*}}  & = &  -2 i \lambda a \left[ \cos(\frac{p_{1}\times p_{2}}{2})
       + \cos( {p_{1}\times p_{3}\over 2}) + \cos( {p_{2}\times p_{3}\over 
       2}) \right] \\ \nonumber \\ \nonumber \\
  \parbox{70pt}{\begin{fmfgraph*}(50,50)
     \fmfsurround{v1,v2,v3}
     \fmfdot{c}
     \fmf{plain}{v1,c}
     \fmf{dashes}{v2,c}
     \fmf{dashes}{v3,c}
     \fmfv{label=$\sigma \hspace{5pt} p_{3}$,label.angle=60}{v1}
     \fmfv{label=$\pi^{*}_{i}   \hspace{5pt} p_{1}$,label.angle=180}{v2}
     \fmfv{label=$\pi_{j}  \hspace{5pt} p_{2}$}{v3}
  \end{fmfgraph*}}  & = &  -2 i \lambda a \delta^{ij} 
  \tn{e}^{-\frac{i}{2} (p_{1}\times p_{2})} \\ \nonumber \\ \nonumber \\
  \parbox{70pt}{\begin{fmfgraph*}(50,50)
     \fmfsurround{v1,v2,v3}
     \fmfdot{c}
     \fmf{plain}{v1,c}
     \fmf{dots}{v2,c}
     \fmf{dots}{v3,c}
     \fmfv{label=$\sigma \hspace{5pt} p_{3}$,label.angle=60}{v1}
     \fmfv{label=$\pi_{0} \hspace{5pt} p_{1}$,label.angle=180}{v2}
     \fmfv{label=$\pi_{0} \hspace{5pt} p_{2}$}{v3}
  \end{fmfgraph*}}  & = &  -2 i \lambda a  \cos(\frac{p_{1}
      \times p_{2}}{2}) \\ \nonumber \\ \nonumber \\ 
 \eea     
\end{fmffile}    

\subsection{Scalar Potential Feynman Rules, $U(2)$ Adjoint}

All momenta flow into the vertices.

\begin{fmffile}{vertices2}
  \bea
  \parbox{70pt}{\begin{fmfgraph*}(50,50)
     \fmfsurroundn{v}{8}
     \fmfdot{c}
     \fmf{dashes}{v2,c}
     \fmf{dashes}{v4,c}
     \fmf{dashes}{v6,c}
     \fmf{dashes}{v8,c} 
     \fmfv{label=$\pi \hspace{5pt} p_{2}$}{v2}
     \fmfv{label=$\pi^{*} \hspace{5pt} p_{1}$}{v4}
     \fmfv{label=$\pi^{*} \hspace{5pt} p_{3}$}{v6}
     \fmfv{label=$\pi \hspace{5pt} p_{4}$}{v8}
  \end{fmfgraph*}} & = & - 2 i (\lambda_{1} + \lambda_{2}) 
    \cos(\frac{p_{1} \times p_{2}}{2} + \frac{p_{3} \times p_{4}}{2}) -
    2 i \lambda_{2} \cos(\frac{p_{1} \times p_{3}}{2}) \cos(\frac{p_{2} \times 
    p_{4}}{2})  \nonumber \\ \\ \nonumber \\
  \parbox{70pt}{\begin{fmfgraph*}(50,50)
     \fmfsurroundn{v}{8}
     \fmfdot{c}
     \fmf{dashes}{v2,c}
     \fmf{dashes}{v4,c} 
     \fmf{plain}{v6,c}
     \fmf{plain}{v8,c}
     \fmfv{label=$\pi \hspace{5pt} p_{2}$}{v2}
     \fmfv{label=$\pi^{*} \hspace{5pt} p_{1}$}{v4}
     \fmfv{label=$\sigma \hspace{5pt} p_{3}$}{v6}
     \fmfv{label=$\sigma \hspace{5pt} p_{4}$}{v8}
  \end{fmfgraph*}} & = & - 2 i (\lambda_{1} + \lambda_{2}) 
    \cos(\frac{p_{1} \times p_{2}}{2}) \cos(\frac{p_{3} \times 
    p_{4}}{2}) + i \lambda_{1} \cos(\frac{p_{1} \times p_{3}}{2} + 
    \frac{p_{2} \times p_{4}}{2})  \nonumber \\ \\ \nonumber \\
  \parbox{70pt}{\begin{fmfgraph*}(50,50)
     \fmfsurroundn{v}{8}
     \fmfdot{c}
     \fmf{dashes}{v2,c}
     \fmf{dashes}{v4,c} 
     \fmf{dots}{v6,c}
     \fmf{dots}{v8,c}
     \fmfv{label=$\pi \hspace{5pt} p_{2}$}{v2}
     \fmfv{label=$\pi^{*} \hspace{5pt} p_{1}$}{v4}
     \fmfv{label=$\phi_{4} \hspace{5pt} p_{3}$}{v6}
     \fmfv{label=$\phi_{4} \hspace{5pt} p_{4}$}{v8}
  \end{fmfgraph*}} & = & - 2 i (\lambda_{1} + \lambda_{2}) 
    \cos(\frac{p_{1} \times p_{2}}{2}) \cos(\frac{p_{3} \times 
    p_{4}}{2}) - i \lambda_{1} \cos(\frac{p_{1} \times p_{3}}{2} + 
    \frac{p_{2} \times p_{4}}{2})  \nonumber \\ \\ \nonumber \\
  \parbox{70pt}{\begin{fmfgraph*}(50,50)
     \fmfsurroundn{v}{8}
     \fmfdot{c}
     \fmf{plain}{v2,c}
     \fmf{plain}{v4,c} 
     \fmf{dots}{v6,c}
     \fmf{dots}{v8,c}
     \fmfv{label=$\sigma \hspace{5pt} p_{2}$}{v2}
     \fmfv{label=$\sigma \hspace{5pt} p_{1}$}{v4}
     \fmfv{label=$\phi_{4} \hspace{5pt} p_{3}$}{v6}
     \fmfv{label=$\phi_{4} \hspace{5pt} p_{4}$}{v8}
  \end{fmfgraph*}} & = & - 2 i (\lambda_{1} + \lambda_{2}) 
    \cos(\frac{p_{1} \times p_{2}}{2}) \cos(\frac{p_{3} \times 
    p_{4}}{2}) - i \lambda_{1} \cos(\frac{p_{1} \times p_{3}}{2} + 
    \frac{p_{2} \times p_{4}}{2})  \nonumber \\ \\ \nonumber \\
  \parbox{70pt}{\begin{fmfgraph*}(50,50)
     \fmfsurroundn{v}{8}
     \fmfdot{c}
     \fmf{plain}{v2,c}
     \fmf{plain}{v4,c} 
     \fmf{plain}{v6,c}
     \fmf{plain}{v8,c}
     \fmfv{label=$\sigma(\phi_{4}) \hspace{5pt} p_{2}$}{v2}
     \fmfv{label=$\sigma(\phi_{4}) \hspace{5pt} p_{1}$}{v4}
     \fmfv{label=$\sigma(\phi_{4}) \hspace{5pt} p_{3}$}{v6}
     \fmfv{label=$\sigma(\phi_{4})\hspace{5pt} p_{4}$}{v8}
  \end{fmfgraph*}} & = & - 2 i \lambda \left[ \cos(\frac{p_{1} \times p_{2}}{2})
    \cos(\frac{p_{3} \times p_{4}}{2}) + \cos(\frac{p_{1} \times p_{3}}{2})
    \cos(\frac{p_{2} \times p_{4}}{2}) \right. \nonumber \\ & & \left. + \cos(\frac{p_{1} \times 
    p_{4}}{2}) \cos(\frac{p_{2} \times p_{3}}{2}) \right] \nonumber \\ \\ \nonumber \\
  \parbox{70pt}{\begin{fmfgraph*}(50,50)
     \fmfsurroundn{v}{8}
     \fmfdot{c}
     \fmf{dots}{v2,c}
     \fmf{dashes}{v4,c} 
     \fmf{dashes}{v6,c}
     \fmf{plain}{v8,c}
     \fmfv{label=$\phi_{4} \hspace{5pt} p_{2}$}{v2}
     \fmfv{label=$\pi^{*} \hspace{5pt} p_{1}$}{v4}
     \fmfv{label=$\pi \hspace{5pt} p_{3}$}{v6}
     \fmfv{label=$\sigma \hspace{5pt} p_{4}$}{v8}
  \end{fmfgraph*}} & = & - \lambda_{1} \sin(\frac{p_{1} \times 
      p_{2}}{2}+ \frac{p_{3} \times p_{4}}{2}) \nonumber \\ \\ \nonumber \\
  \parbox{70pt}{\begin{fmfgraph*}(50,50)
     \fmfsurround{v1,v2,v3}
     \fmfdot{c}
     \fmf{plain}{v1,c}
     \fmf{dashes}{v2,c}
     \fmf{dashes}{v3,c}
     \fmfv{label=$\sigma \hspace{5pt} p_{3}$,label.angle=60}{v1}
     \fmfv{label=$\pi^{*}  \hspace{5pt} p_{1}$,label.angle=180}{v2}
     \fmfv{label=$\pi \hspace{5pt} p_{2}$}{v3}
  \end{fmfgraph*}}  & = &  -2 i \lambda a \cos(\frac{p_{1}\times 
  p_{2}}{2}) \\ \nonumber \\ \nonumber \\
  \parbox{70pt}{\begin{fmfgraph*}(50,50)
     \fmfsurround{v1,v2,v3}
     \fmfdot{c}
     \fmf{plain}{v1,c}
     \fmf{plain}{v2,c}
     \fmf{plain}{v3,c}
     \fmfv{label=$\sigma \hspace{5pt} p_{3}$,label.angle=60}{v1}
     \fmfv{label=$\sigma \hspace{5pt} p_{1}$,label.angle=180}{v2}
     \fmfv{label=$\sigma \hspace{5pt} p_{2}$}{v3}
  \end{fmfgraph*}}  & = &  -2 i \lambda a \left[ \cos(\frac{p_{1}\times p_{2}}{2})
       + \cos( {p_{1}\times p_{3}\over 2}) + \cos( {p_{2}\times p_{3}\over 
       2}) \right] \\ \nonumber \\ \nonumber \\
  \parbox{70pt}{\begin{fmfgraph*}(50,50)
     \fmfsurround{v1,v2,v3}
     \fmfdot{c}
     \fmf{plain}{v1,c}
     \fmf{dots}{v2,c}
     \fmf{dots}{v3,c}
     \fmfv{label=$\sigma \hspace{5pt} p_{3}$,label.angle=60}{v1}
     \fmfv{label=$\phi_{4} \hspace{5pt} p_{1}$,label.angle=180}{v2}
     \fmfv{label=$\phi_{4} \hspace{5pt} p_{2}$}{v3}
  \end{fmfgraph*}}  & = &  -2 i (\lambda + \lambda_{1})  a  \cos(\frac{p_{1}
      \times p_{2}}{2}) \\ \nonumber \\ \nonumber \\
  \parbox{70pt}{\begin{fmfgraph*}(50,50)
     \fmfsurround{v1,v2,v3}
     \fmfdot{c}
     \fmf{dots}{v1,c}
     \fmf{dashes}{v2,c}
     \fmf{dashes}{v3,c}
     \fmfv{label=$\phi_{4} \hspace{5pt} p_{3}$,label.angle=60}{v1}
     \fmfv{label=$\pi \hspace{5pt} p_{1}$,label.angle=180}{v2}
     \fmfv{label=$\pi^{*} \hspace{5pt} p_{2}$}{v3}
  \end{fmfgraph*}}  & = &  - \lambda_{1} a  \sin(\frac{p_{1}
      \times p_{2}}{2})  
 \eea
\end{fmffile}

\subsection{Scalar Potential Feynman Rules, O(4) Adjoint (Partial)}

All momenta flow into the vertices.

\begin{fmffile}{vertices3}
  \bea
  \parbox{70pt}{\begin{fmfgraph*}(50,50)
     \fmfsurroundn{v}{8}
     \fmfdot{c}
     \fmf{dashes}{v2,c}
     \fmf{dashes}{v4,c}
     \fmf{dashes}{v6,c}
     \fmf{dashes}{v8,c} 
     \fmfv{label=$\pi_{1} \hspace{5pt} p_{2}$}{v2}
     \fmfv{label=$\pi_{1} \hspace{5pt} p_{1}$}{v4}
     \fmfv{label=$\pi_{1} \hspace{5pt} p_{3}$}{v6}
     \fmfv{label=$\pi_{1} \hspace{5pt} p_{4}$}{v8}
  \end{fmfgraph*}} & = & - 2 i (\lambda_{1} + \frac{\lambda_{2}}{4}) 
     \left[ \cos(\frac{p_{1} \times p_{2}}{2})
    \cos(\frac{p_{3} \times p_{4}}{2}) + \cos(\frac{p_{1} \times p_{3}}{2})
    \cos(\frac{p_{2} \times p_{4}}{2}) \right. \nonumber \\ & & \left. + 
    \cos(\frac{p_{1} \times 
    p_{4}}{2}) \cos(\frac{p_{2} \times p_{3}}{2}) \right]
 \nonumber \\ \\ \nonumber \\
   \parbox{70pt}{\begin{fmfgraph*}(50,50)
     \fmfsurroundn{v}{8}
     \fmfdot{c}
     \fmf{dashes}{v2,c}
     \fmf{dashes}{v4,c}
     \fmf{dashes}{v6,c}
     \fmf{dashes}{v8,c} 
     \fmfv{label=$\pi_{1} \hspace{5pt} p_{2}$}{v2}
     \fmfv{label=$\pi_{1} \hspace{5pt} p_{1}$}{v4}
     \fmfv{label=$\pi_{2} \hspace{5pt} p_{3}$}{v6}
     \fmfv{label=$\pi_{2} \hspace{5pt} p_{4}$}{v8}
  \end{fmfgraph*}} & = & - 2 i (\lambda_{1} + \frac{\lambda_{2}}{2})
      \cos(\frac{p_{1}\times p_{2}}{2}) \cos(\frac{p_{3}\times p_{4}}{2})
      + \frac{i\lambda_{2}}{2} \cos(\frac{p_{1}\times p_{3}}{2}+\frac{p_{2}
      \times p_{4}}{2})  \nonumber \\ \\ \nonumber \\
   \parbox{70pt}{\begin{fmfgraph*}(50,50)
     \fmfsurroundn{v}{8}
     \fmfdot{c}
     \fmf{dashes}{v2,c}
     \fmf{dashes}{v4,c}
     \fmf{plain}{v6,c}
     \fmf{plain}{v8,c} 
     \fmfv{label=$\pi_{1} \hspace{5pt} p_{2}$}{v2}
     \fmfv{label=$\pi_{1} \hspace{5pt} p_{1}$}{v4}
     \fmfv{label=$\alpha(\beta)(\psi) \hspace{5pt} p_{3}$}{v6}
     \fmfv{label=$\alpha(\beta)(\psi) \hspace{5pt} p_{4}$}{v8}
  \end{fmfgraph*}} & = & - 2 i (\lambda_{1} + \frac{\lambda_{2}}{2})
      \cos(\frac{p_{1}\times p_{2}}{2}) \cos(\frac{p_{3}\times p_{4}}{2})
      - \frac{i\lambda_{2}}{2} \cos(\frac{p_{1}\times p_{3}}{2}+\frac{p_{2}
      \times p_{4}}{2})  \nonumber \\ \\ \nonumber \\
   \parbox{70pt}{\begin{fmfgraph*}(50,50)
     \fmfsurroundn{v}{8}
     \fmfdot{c}
     \fmf{dashes}{v2,c}
     \fmf{dashes}{v4,c}
     \fmf{plain}{v6,c}
     \fmf{plain}{v8,c} 
     \fmfv{label=$\pi_{1} \hspace{5pt} p_{2}$}{v2}
     \fmfv{label=$\pi_{1} \hspace{5pt} p_{1}$}{v4}
     \fmfv{label=$\sigma \hspace{5pt} p_{3}$}{v6}
     \fmfv{label=$\sigma \hspace{5pt} p_{4}$}{v8}
  \end{fmfgraph*}} & = & - 2 i (\lambda_{1} + \frac{\lambda_{2}}{2})
      \cos(\frac{p_{1}\times p_{2}}{2}) \cos(\frac{p_{3}\times p_{4}}{2})
      + \frac{i\lambda_{2}}{2} \cos(\frac{p_{1}\times p_{3}}{2}+\frac{p_{2}
      \times p_{4}}{2})  \nonumber \\ \\ \nonumber \\
  \parbox{70pt}{\begin{fmfgraph*}(50,50)
     \fmfsurround{v1,v2,v3}
     \fmfdot{c}
     \fmf{plain}{v1,c}
     \fmf{dashes}{v2,c}
     \fmf{dashes}{v3,c}
     \fmfv{label=$\sigma \hspace{5pt} p_{3}$,label.angle=60}{v1}
     \fmfv{label=$\pi_{1(2)} \hspace{5pt} p_{1}$,label.angle=180}{v2}
     \fmfv{label=$\pi_{1(2)} \hspace{5pt} p_{2}$}{v3}
  \end{fmfgraph*}}  & = &  -2 i (\lambda_{1} + \frac{\lambda_{2}}{4}) 
    a \cos(\frac{p_{1}\times p_{2}}{2}) \\ \nonumber \\ \nonumber \\
  \parbox{70pt}{\begin{fmfgraph*}(50,50)
     \fmfsurround{v1,v2,v3}
     \fmfdot{c}
     \fmf{plain}{v1,c}
     \fmf{plain}{v2,c}
     \fmf{plain}{v3,c}
     \fmfv{label=$\sigma \hspace{5pt} p_{3}$,label.angle=60}{v1}
     \fmfv{label=$\sigma \hspace{5pt} p_{1}$,label.angle=180}{v2}
     \fmfv{label=$\sigma \hspace{5pt} p_{2}$}{v3}
  \end{fmfgraph*}}  & = &  -2 i (\lambda_{1}+\frac{\lambda_{2}}{4}) a 
     \left[ \cos(\frac{p_{1}\times p_{2}}{2})
      + \cos( {p_{1}\times p_{3}\over 2}) + \cos( {p_{2}\times p_{3}\over 
       2}) \right] \\ \nonumber \\ \nonumber \\
  \parbox{70pt}{\begin{fmfgraph*}(50,50)
     \fmfsurround{v1,v2,v3}
     \fmfdot{c}
     \fmf{plain}{v1,c}
     \fmf{plain}{v2,c}
     \fmf{plain}{v3,c}
     \fmfv{label=$\sigma \hspace{5pt} p_{3}$,label.angle=60}{v1}
     \fmfv{label=$\alpha(\beta)(\psi) \hspace{5pt} p_{1}$,label.angle=180}{v2}
     \fmfv{label=$\alpha(\beta)(\psi) \hspace{5pt} p_{2}$}{v3}
  \end{fmfgraph*}}  & = & -2 i (\lambda_{1} + \frac{3 \lambda_{2}}{4}) 
    a \cos(\frac{p_{1}\times p_{2}}{2}) \\ \nonumber \\ \nonumber \\
\eea

\end{fmffile}


\end{document}